\documentclass{article}

\title{Quantum algorithm for the classification of supersymmetric top quark events}

\author{Pedrame Bargassa$^{1,2}$, Timoth\'ee Cabos$^3$, Samuele Cavinato$^{4,5}$, \\
Artur Cordeiro Oudot Choi$^3$, Timoth\'ee Hessel$^3$}
\date{
  $^1$Physics of Information and Quantum Technologies Group, Instituto de
  Telecomunica\c{c}\~{o}es, Lisbon, Portugal\\
  $^2$Laborat\'orio de Instrumenta\c{c}\~ao e F\'isica Experimental de Part\'iculas, Lisbon, Portugal\\
  $^3$Sorbonne Universit\'e, Paris, France\\
  $^4$Dipartimento di Fisica e Astronomia ``G. Galilei'', Universit\`a di Padova, Italy\\
  $^5$Medical Physics Deparment, Veneto Institute of Oncology IOV-IRCCS, Padova, Italy \\[2ex]
}
\usepackage{amsfonts}
\usepackage{graphicx}
\usepackage{lineno}
\usepackage{xcolor}
\usepackage[colorlinks = true,
            linkcolor = blue,
            urlcolor  = blue,
            citecolor = blue,
            anchorcolor = blue]{hyperref}
\newcommand{\pt}{\ensuremath{p_{\mathrm{T}}}~}
\newcommand{\ptl}{\ensuremath{p_{\mathrm{T}}(l)}}
\newcommand{\etl}{\ensuremath{\eta(l)}~}
\newcommand{\chl}{\ensuremath{Q(l)}~}
\newcommand{\met}{\ensuremath{E_\mathrm{T}^\mathrm{miss}}}
\newcommand{\mt}{\ensuremath{M_{\mathrm{T}}}}
\newcommand{\Ht}{\ensuremath{H_{\mathrm{T}}}}
\newcommand{\jeti}[1]{\ensuremath{j_{#1}}}
\newcommand{\ptisr}{\ensuremath{p_{\mathrm{T}}(\jeti{1})}~}
\newcommand{\njet}{\ensuremath{N(\mathrm{jets})}~}
\newcommand{\nbl}{\ensuremath{N(b)}~}
\newcommand{\ptb}{\ensuremath{p_{\mathrm{T}}(b)}~}
\newcommand{\bdisc}{\ensuremath{\mathrm{Disc(b)}}~}
\newcommand{\drLB}{\ensuremath{\Delta R (l, b)}~}
\newcommand{\wjets}{\ensuremath{W+}jets~}
\newcommand{\ttbar}{\ensuremath{\mathrm{t}\bar{\mathrm{t}}}~}
\newcommand{\stp}{\ensuremath{\tilde{t}_{1}}~}
\newcommand{\lsp}{\ensuremath{\tilde{\chi}_{1}^{0}}~}

\begin{document}

\maketitle

\begin{abstract}

The search for supersymmetric particles is one of the major goals of the 
Large Hadron Collider (LHC). Supersymmetric top (stop) searches play a 
very important role in this respect, but the unprecedented collision rate 
to be attained at the next high luminosity phase of the LHC poses new 
challenges for the separation between any new signal and the standard 
model background. The massive parallelism provided by quantum computing 
techniques may yield an efficient solution for this problem. In this paper 
we show a novel application of the zoomed quantum annealing machine 
learning approach to classify the stop signal versus the background, and 
implement it in a quantum annealer machine. We show that this approach 
together with the preprocessing of the data with principal component 
analysis may yield better results than conventional multivariate 
approaches.

\end{abstract}

\section{Introduction}
\label{s:intro}

After attaining its nominal energy, the Large Hadron Collider (LHC) will 
reach an unprecedented collision rate during its high luminosity phase, 
opening the stage to discoveries beyond the standard model (SM) of 
particle physics. One of the most challenging tasks in searches taking 
place at the LHC is the capacity to categorize events of new phenomena 
(signal) and those of SM processes (background) which mimic the signal. 
Machine learning (ML) tools are among the most powerful means for 
separating signal from background events, having been key to the discovery 
of, e.g., the Higgs boson \cite{HgDisc1,HgDisc2}. More recently, 
quantum annealing for machine learning (QAML) \cite{nature} and its 
zooming variant (QAML-Z) \cite{qamlz} represent the first examples of a 
quantum approach to a classification problem in high energy physics (HEP).

In this paper, we study the application of the QAML-Z algorithm to
the selection of supersymmetric top quark (stop) versus SM events. It is
important to test this algorithm on a new classification problem where
both the abundance of signal versus background events, and their overlap
in the experimental observables are different from~\cite{qamlz},
therefore allowing us to have a better assessment of its classification
capability.
A result on the stop search based on the data accumulated by the LHC in 
2016 (35.9 fb$^{-1}$) has been published~\cite{st4bd}. It is based on a 
classical ML tool which will serve as a reference for gauging the 
performance of the new classifiers. The variables discriminating between 
the stop signal and the SM background, which are used to train the QAML-Z 
algorithm, are based on the same ones as in the classical ML tool 
\cite{st4bd}. We present results of the QAML-Z algorithm for different 
schemes of zoomed quantum annealing, and various sets of variables used in 
the annealer. Also, we introduce a preprocessing of the data through a 
principal component analysis~\cite{pca} (PCA) before feeding it to the 
annealer.

\section{Search for supersymmetric top quark}
\label{s:stop}

\begin{figure}[!htbp]
\begin{center}
\includegraphics{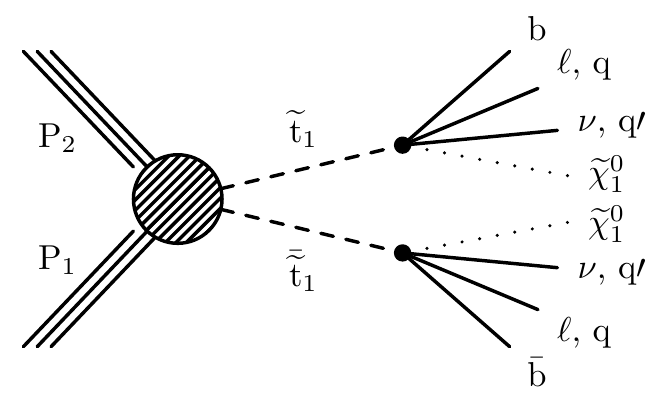} 
\end{center}
\caption{Stop pair production at the LHC with four-body decays.}
\label{fig:4bod}
\end{figure}

One of the main objectives of the physics program at the LHC are searches 
for supersymmetry (SUSY)~\cite{SUSY0,SUSY1,SUSY2,SUSY3,SUSY4,Martin}, one 
of the most promising extensions of the SM. SUSY predicts superpartners of 
SM particles (sparticles) having the same gauge interactions, and whose 
spin differs by one-half unit with respect to their SM partners. The 
search for SUSY has special interest in view of the recent discovery of a 
Higgs boson \cite{HgDisc1,HgDisc2} as it naturally solves the problem of 
quadratically divergent loop corrections to the Higgs boson mass. In this 
study we describe the classification aspect of a search for the pair 
production of the lightest supersymmetric partner of the top quark \stp at 
the LHC machine at $\sqrt{s}$ = 13 TeV, where each stop decays in four 
objects, see Fig.~\ref{fig:4bod}. The lightest neutralino \lsp is 
considered to be stable as the lightest supersymmetric particle. The final 
states considered contain jets, missing transverse energy (\met), and a 
lepton which can be either a muon or an electron.

\begin{figure}[!htbp]
\begin{center}
\includegraphics[scale=0.21]{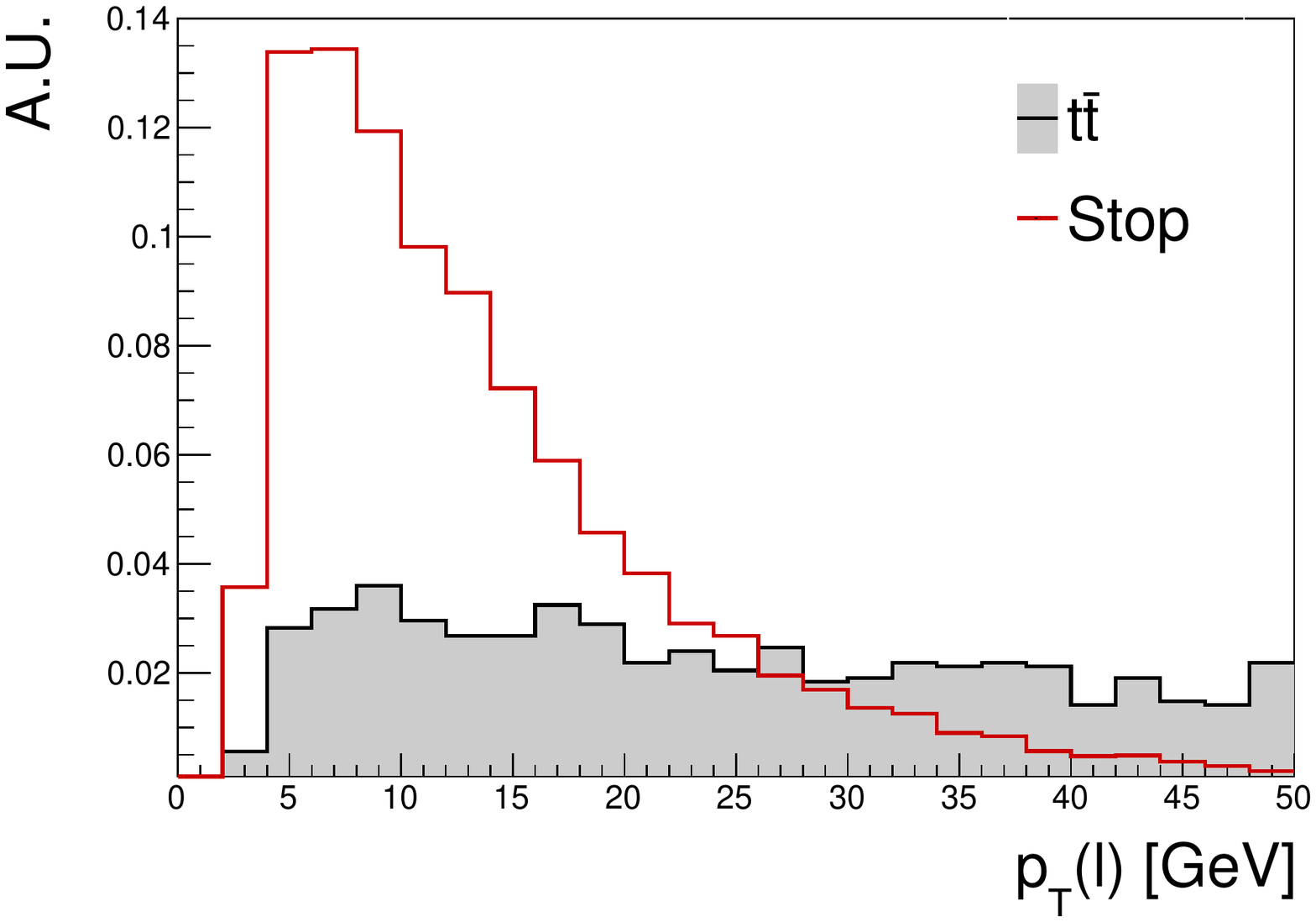}
\includegraphics[scale=0.21]{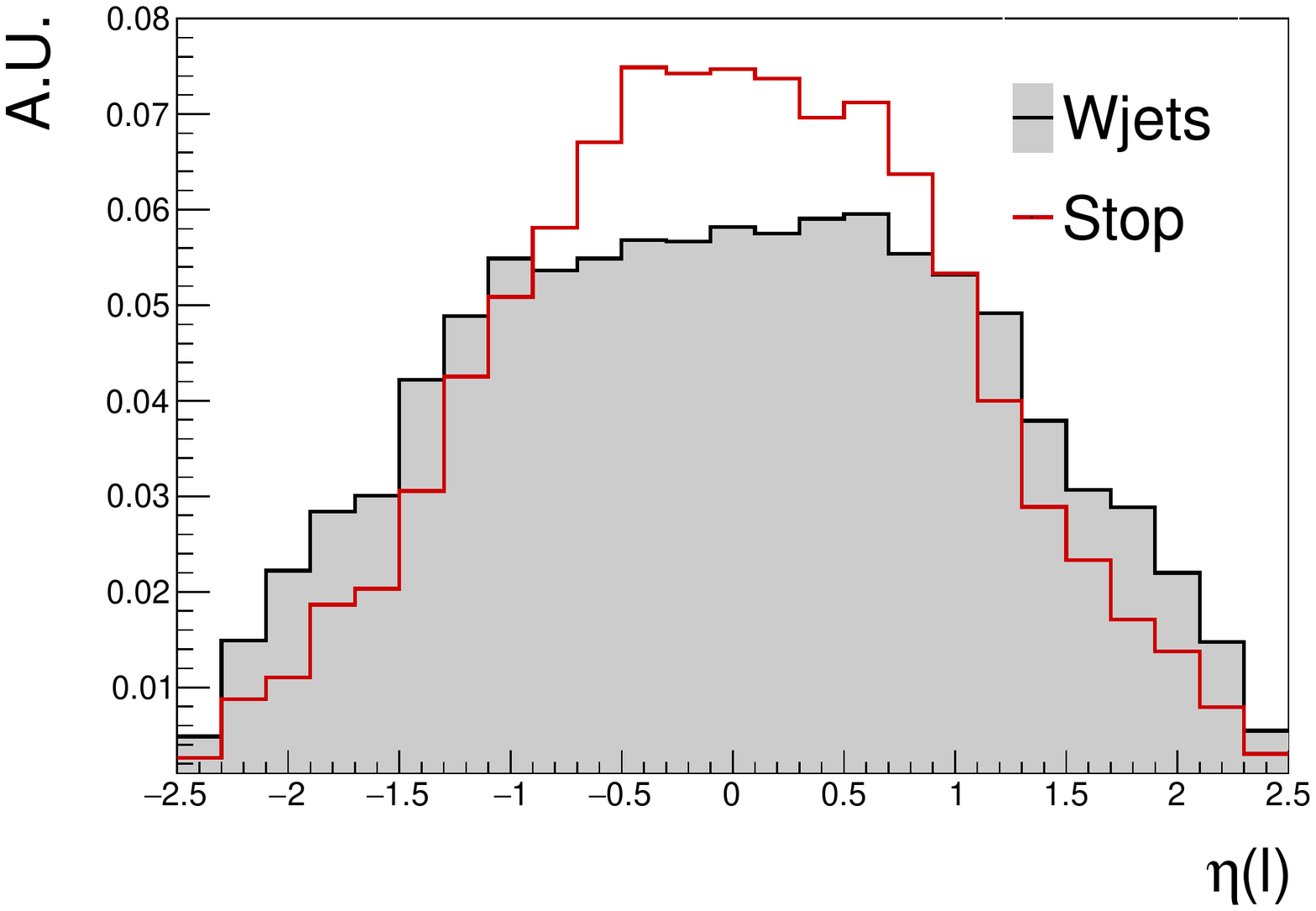}
\includegraphics[scale=0.21]{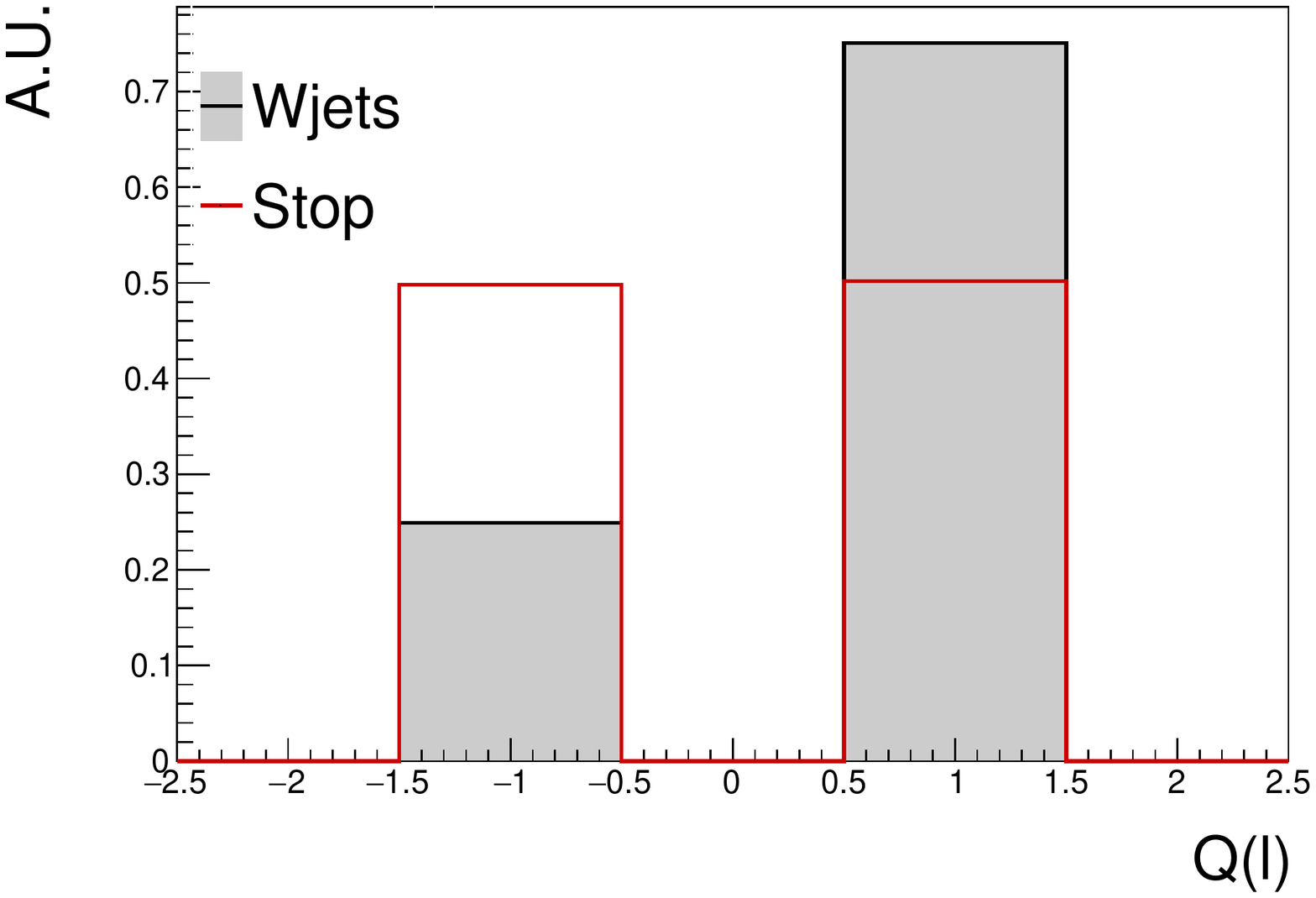} \\
\includegraphics[scale=0.21]{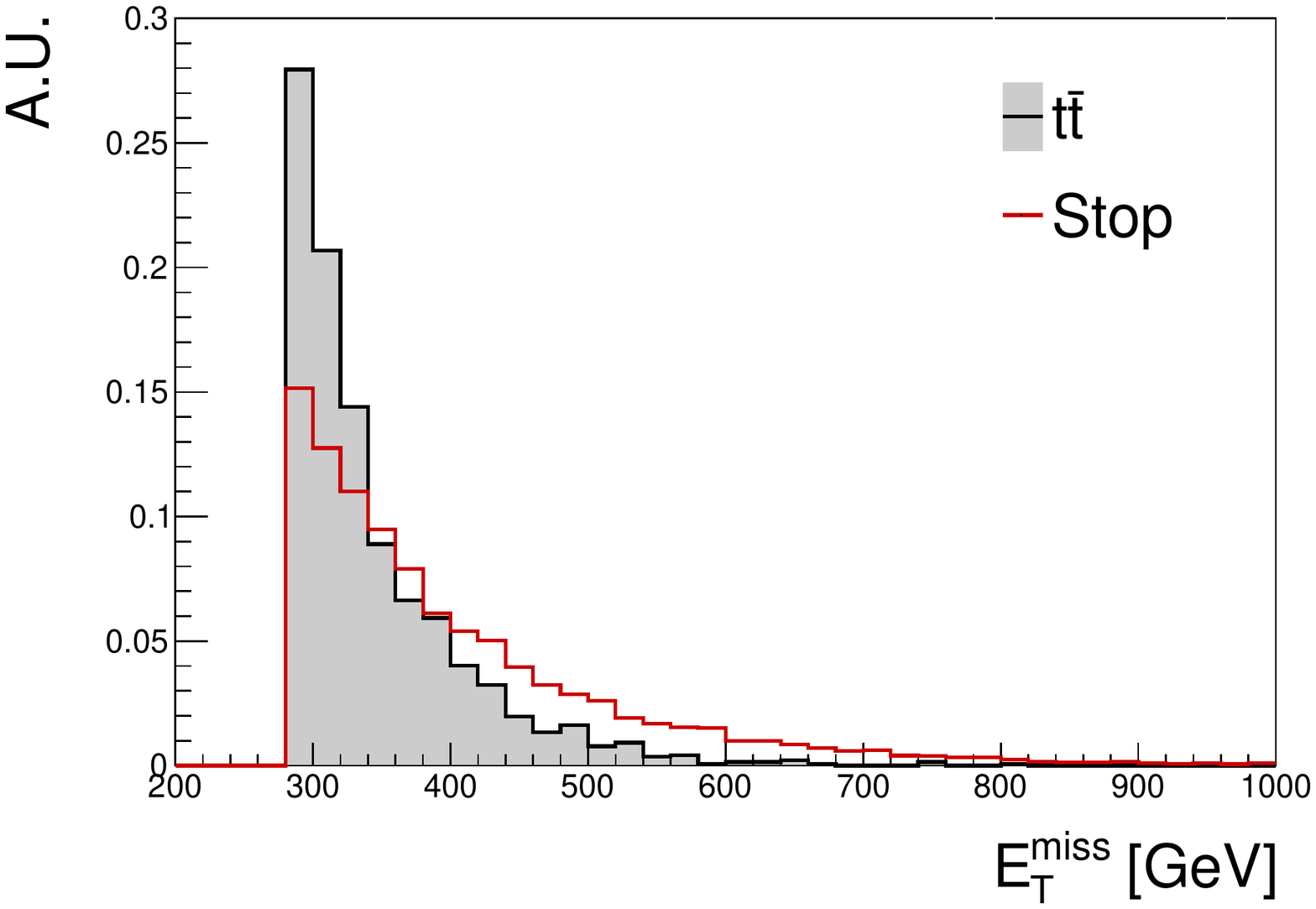}
\includegraphics[scale=0.21]{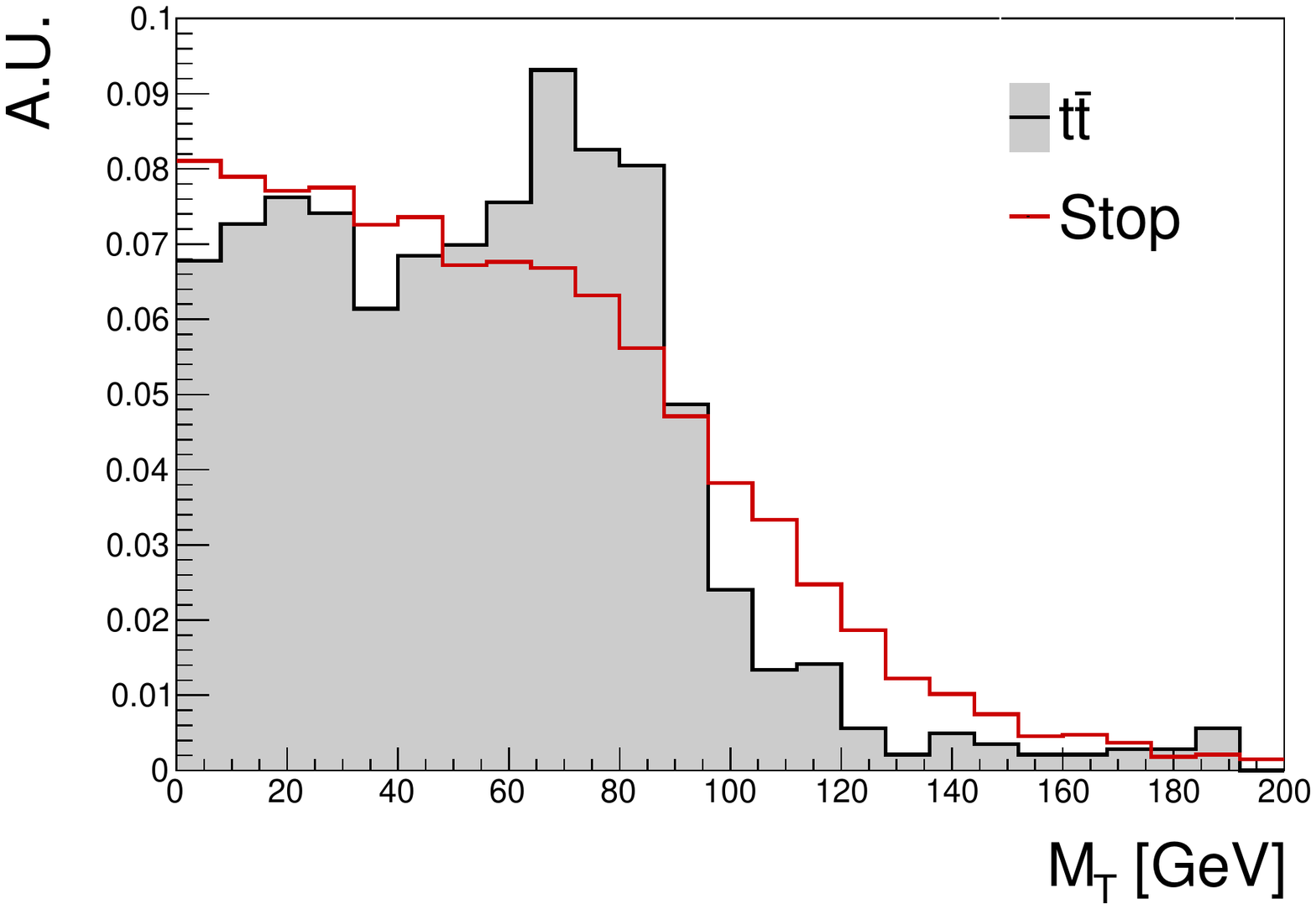}
\includegraphics[scale=0.21]{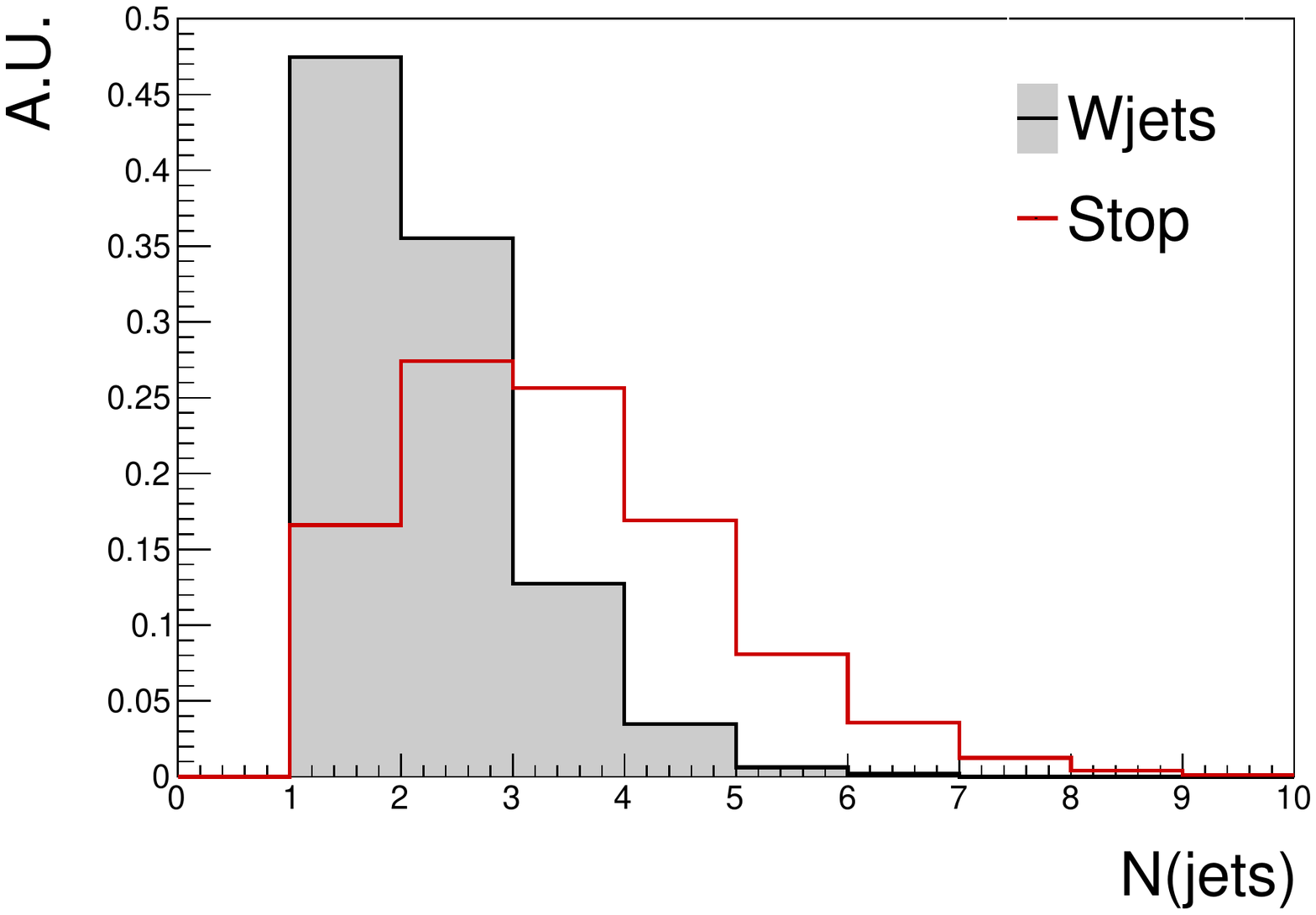} \\
\includegraphics[scale=0.21]{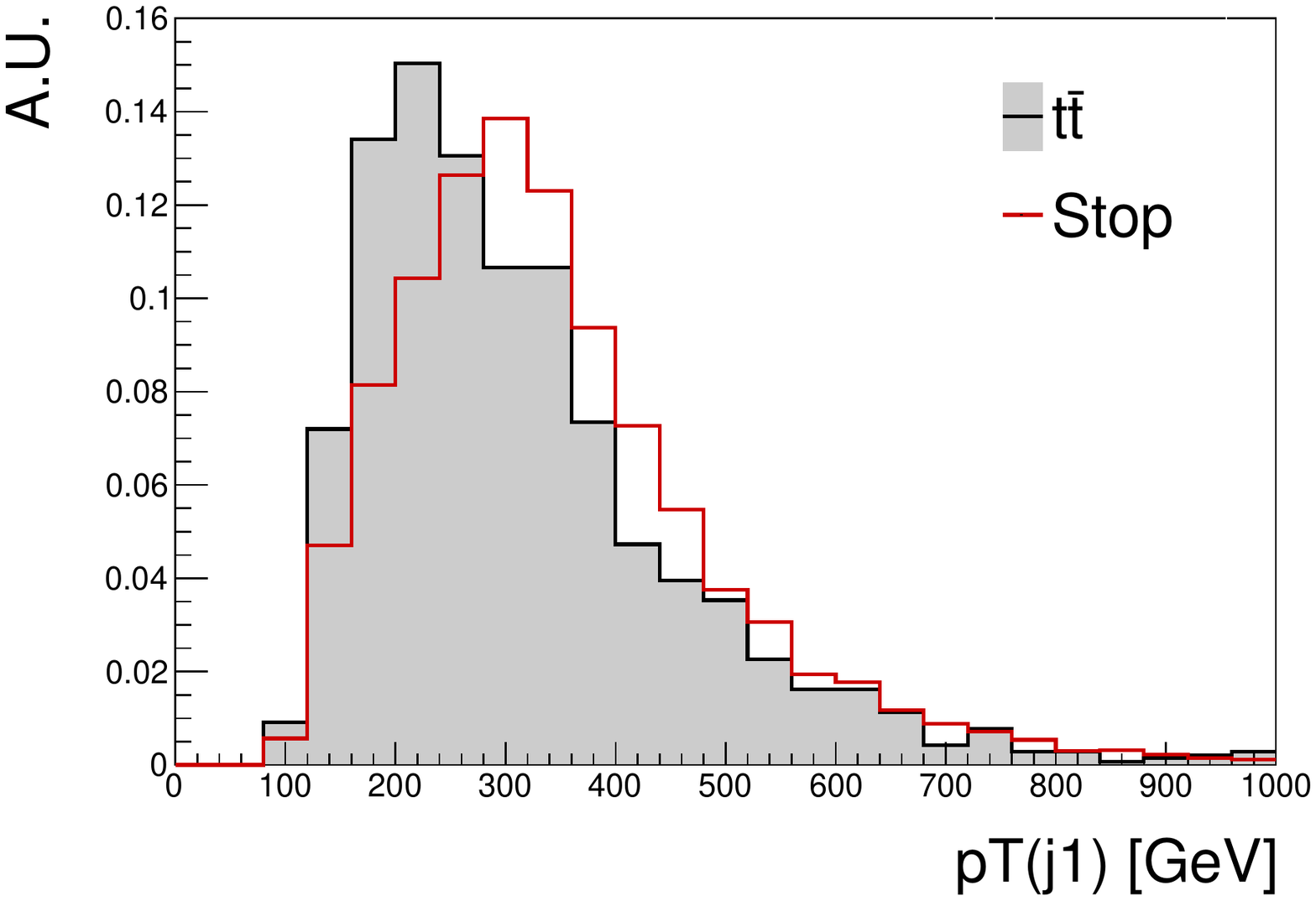}
\includegraphics[scale=0.21]{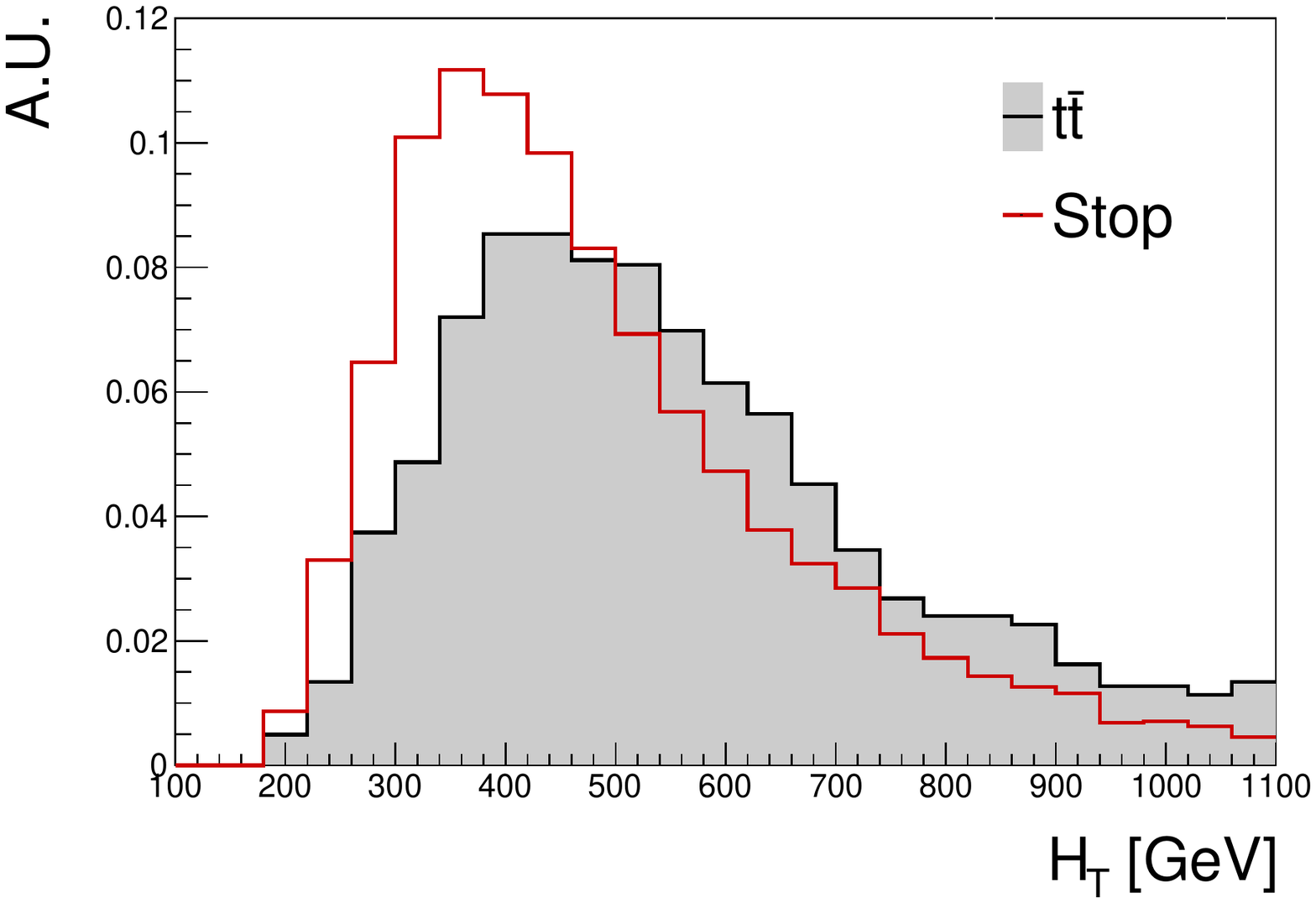}
\includegraphics[scale=0.21]{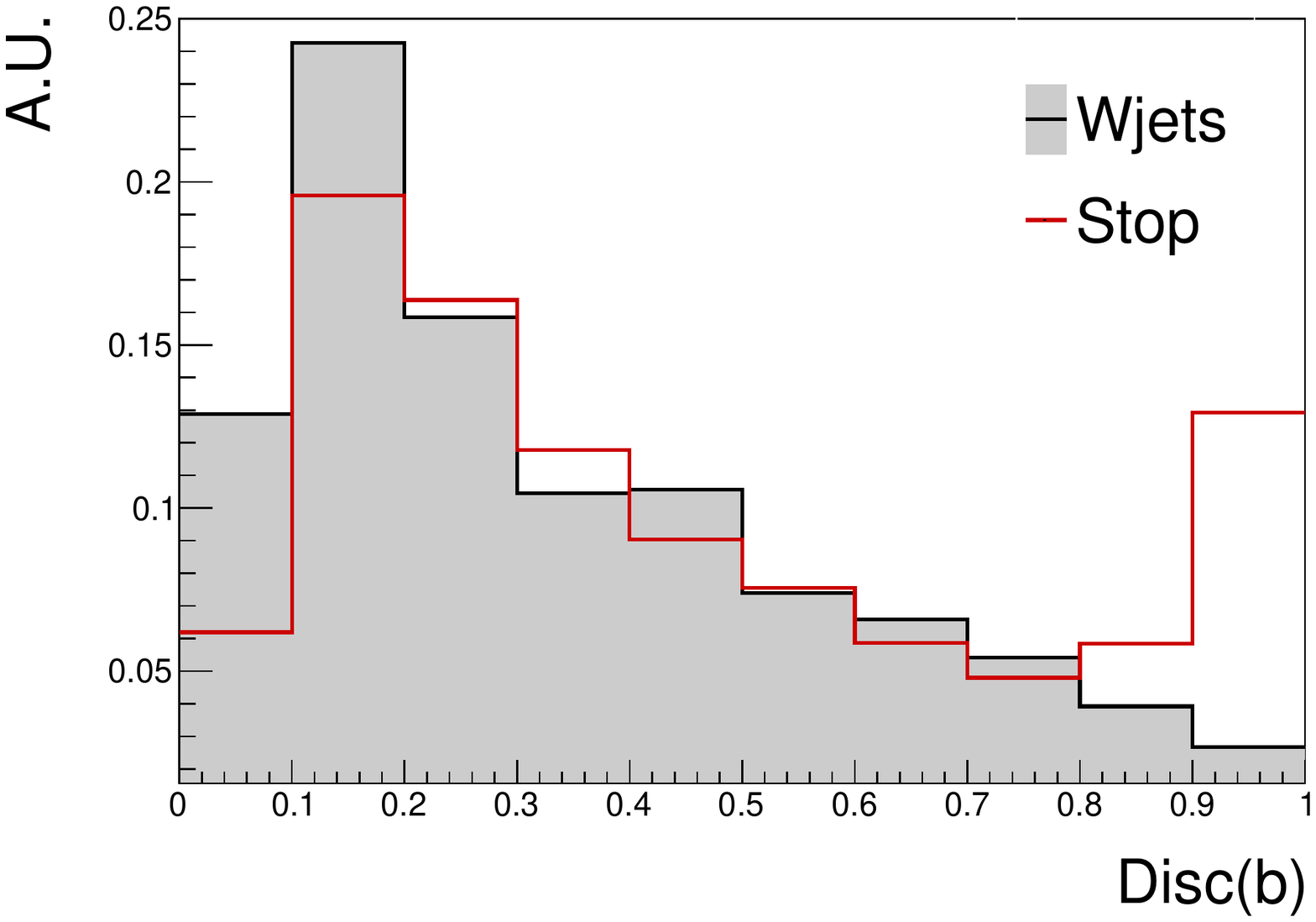} \\
\includegraphics[scale=0.21]{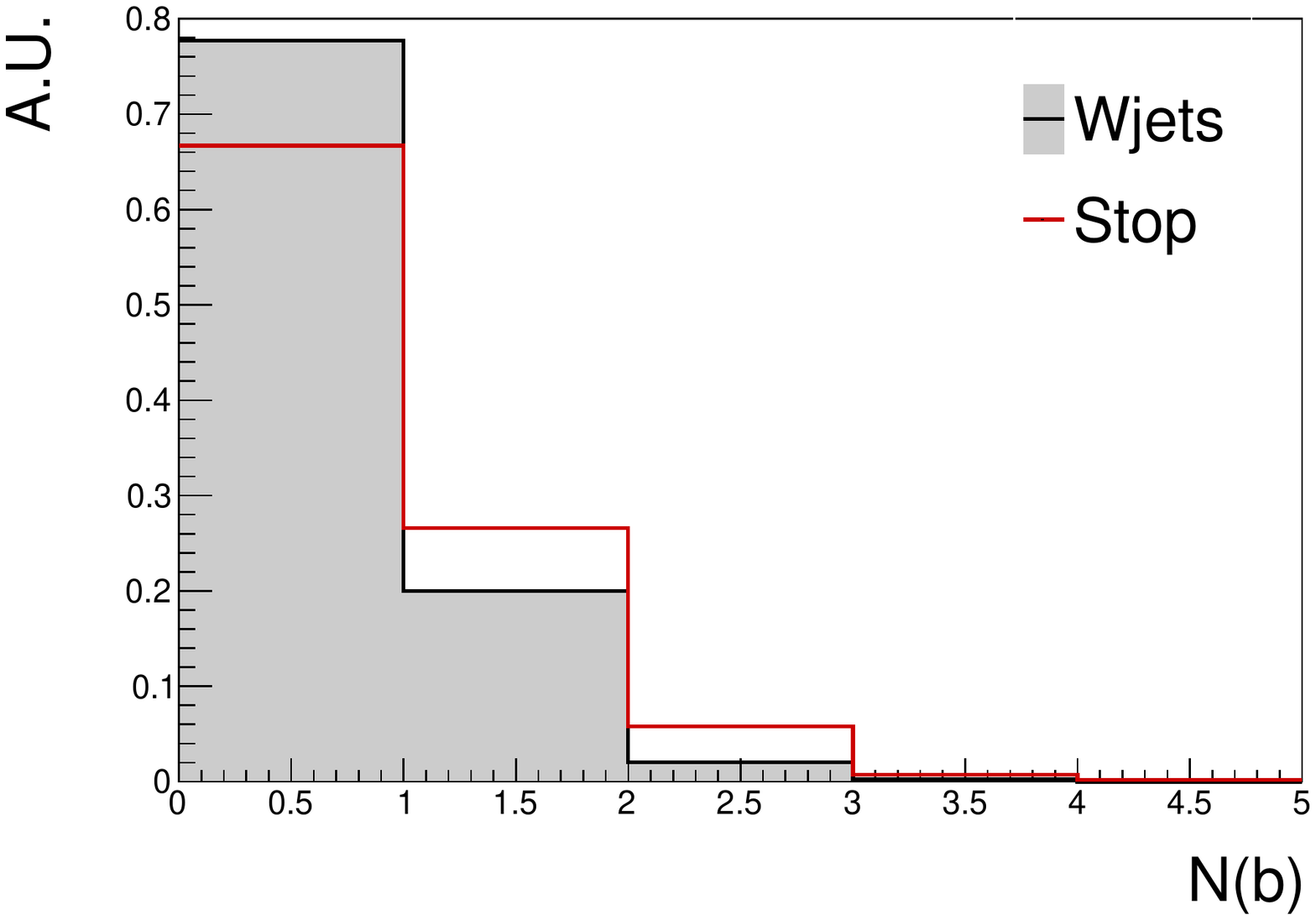}
\includegraphics[scale=0.21]{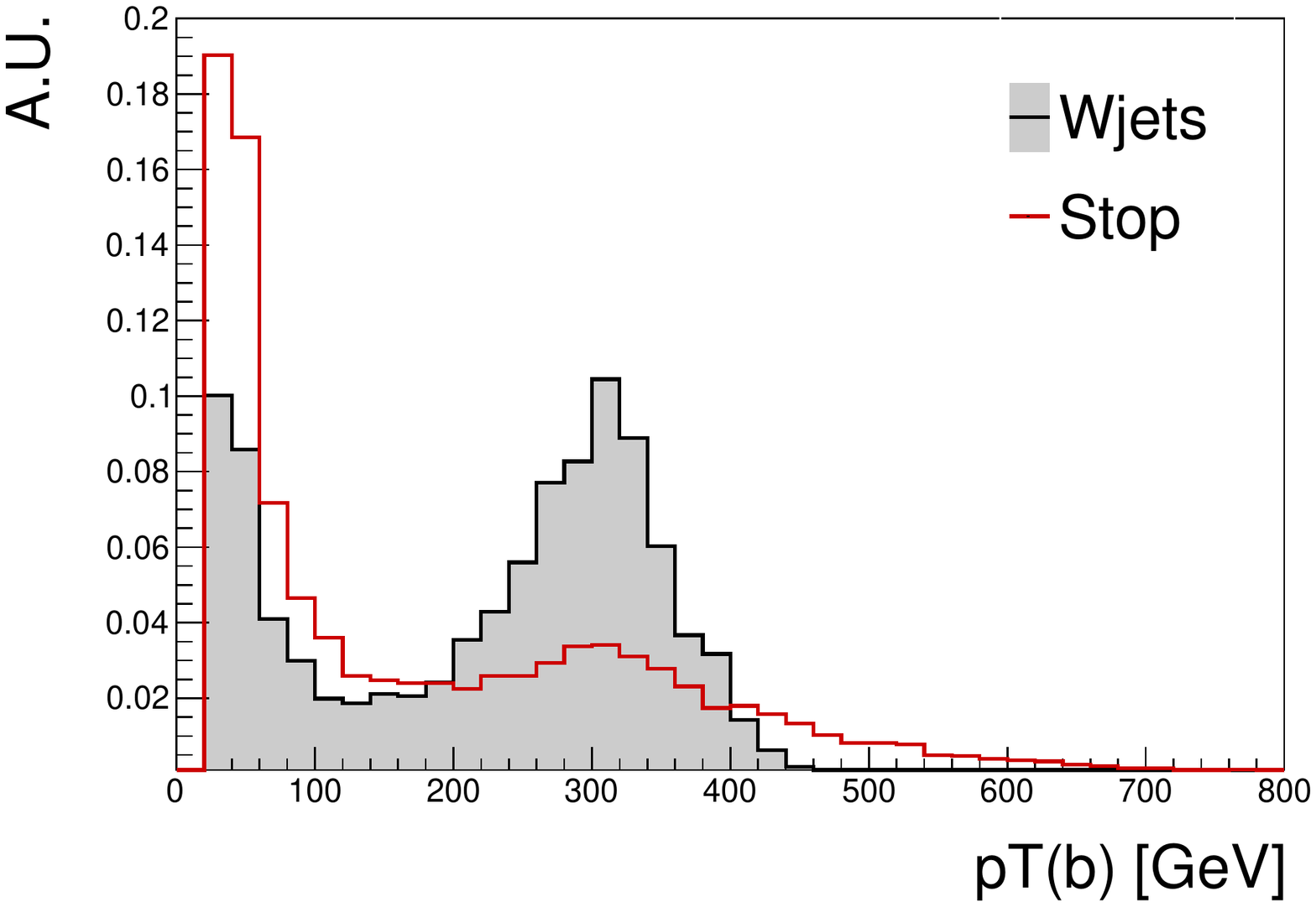}
\includegraphics[scale=0.21]{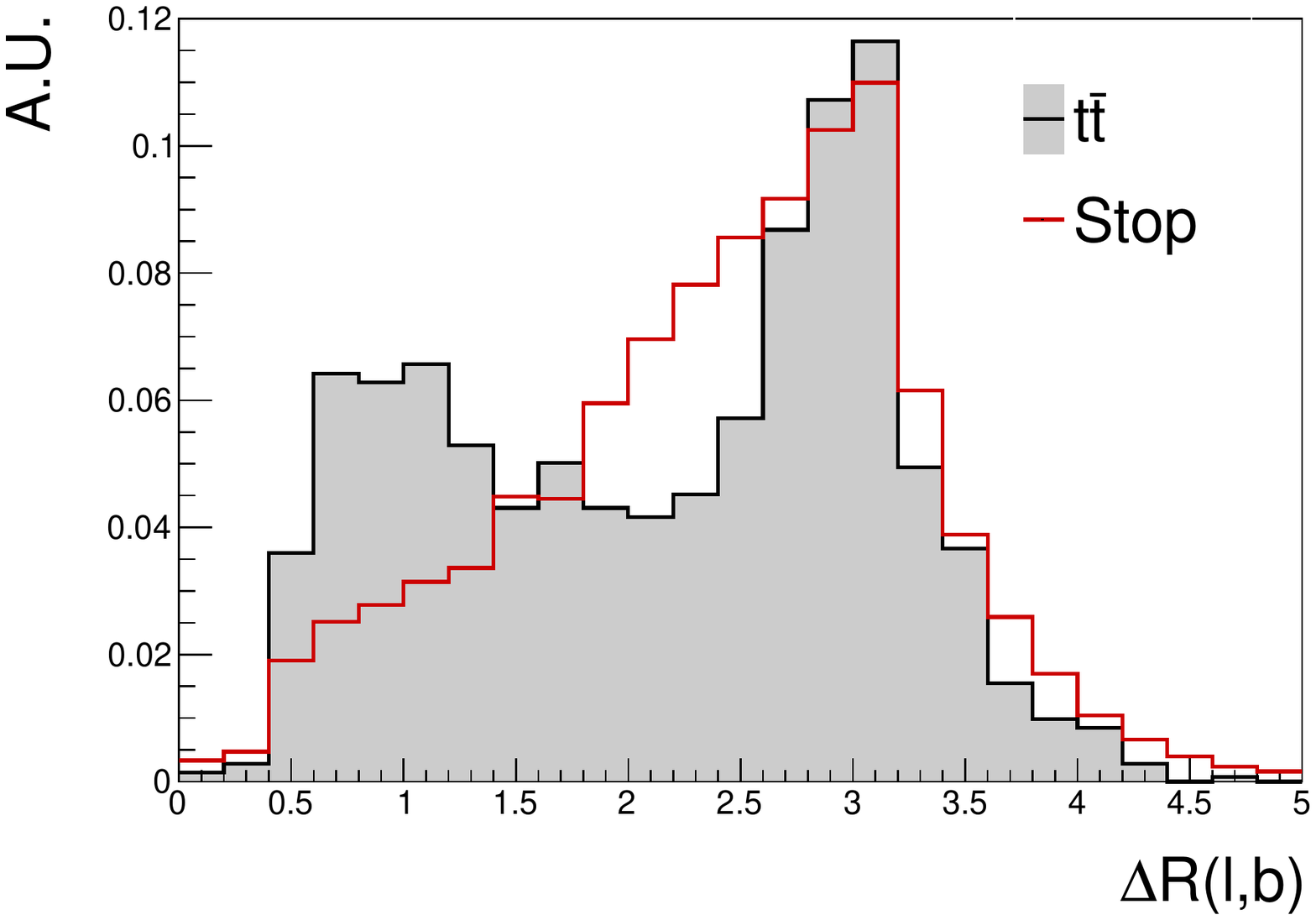}
\end{center}
\caption{Distribution of the discriminating variables for the stop
  signal with $\Delta m = 30$, \wjets and \ttbar, used as input to a BDT in 
  \cite{st4bd}. From top-left to bottom-right: \ptl, \etl, \chl, \met, \mt,
  \njet, \ptisr, \Ht, \bdisc, \nbl, \ptb, \drLB. Distributions are normalized
  to the same area and shown at preselection.}
\label{fig:vardm}
\end{figure}

The sensitivity of this type of search is dominated by the capacity to 
distinguish the stop signal from background events, whose production 
dominates the signal by several orders of magnitude, and whose observables 
overlap with the ones of the stop signal. In this search, the main 
background processes are the \ttbar and \wjets productions. In the search 
based on a classical ML tool~\cite{st4bd}, a preselection is first 
applied to decrease the overwhelming background stemming from the SM. In a 
second stage, boosted decision trees (BDTs)~\cite{RefBDT,tmva} are used to 
classify events as signal and background. To find which variables are the 
most discriminating and should be fed as input to the BDT, different sets 
of variables are tested as input to a BDT whose output is used to maximize 
a figure of merit (FOM)~\cite{fom}:
\begin{eqnarray}
  \mathrm{FOM} & = & \sqrt{ 2 \Big( (S+B)\ln\Big[\frac{(S+B)\cdot(B+\sigma_B^2)}{B^2 + (S+B)\cdot\sigma_B^2}\Big] 
    - \frac{B^2}{\sigma_B^2}\ln\Big[1 + \frac{\sigma_B^2 \cdot S}{B \cdot (B+\sigma_B^2)}\Big] \Big) },
\label{eq:fom}
\end{eqnarray}
where $S$ and $B$ respectively stand for the expected signal and 
background yields for an integrated luminosity of 35.9 fb$^{-1}$ at the 
LHC. The term $\sigma_B = (f \cdot B)$ represents the expected systematic 
uncertainty on the background with $f$ being the relative uncertainty of 
the background yield, taken to be $f=20\%$ as in~\cite{st4bd}. The set of 
variables which maximize the FOM is chosen as the final set of input
variables to the BDT. This metric captures the full information of the 
statistical and systematic uncertainties of a given selection, as it is
important to account for the actual conditions of a search.
The approach based on the maximization of this FOM has been very effective
to find the smallest and most efficient set of discriminating variables
in several searches~\cite{st4bd,st8tev}. A description of the BDT parameters
as well as its development with a FOM maximization procedure as used
in~\cite{st4bd} are provided in Appendix~\ref{s:pb16}. The list of variables
is presented in Table~\ref{tab:vars} and their distribution for signal and
background is provided in Fig.~\ref{fig:vardm}. To render the results of
the classification based on the QAML-Z algorithm as comparable as possible
to the one of~\cite{st4bd}, we use the same preselection of events before
training (see Appendix \ref{s:pb16}). Furthermore, since the FOM as defined
in Eq.~\ref{eq:fom} represents a complete and single-number measure of the
power of a selection, we evaluate the performance of the QAML-Z algorithm
by a maximization of the FOM versus a cut on its output. Finally, for the
comparison of performances to reflect only the difference of a quantum based
versus a classical tool, we train the QAML-Z algorithm with different sets 
made of the same discriminating variables as in the BDT based search 
\cite{st4bd} (see Table~\ref{tab:vars}).

\begin{table}[!htbp]
\begin{center}  
\begin{tabular}{|l|l|}
\hline
\hline
Variable & Description \\
\hline
 \ptl & \pt of the lepton $l$ \\
 \etl & Pseudorapidity of the lepton $l$ \\
 \chl & Charge of the lepton $l$ \\
 \met & Missing transverse energy \\
 \mt  & Transverse invariant mass of \\
      & the (\met,\ptl) system \\
 \njet & Multiplicity of selected jets \\
 \ptisr & \pt of the leading jet \\
 \Ht & $\sum_i p_\mathrm{T}(jet(i))$ \\
 \bdisc & Maximum b-quark tagging discriminant of the jets \\
 \nbl & Number of b-tagged jets \\
 \ptb & \pt of the jet with the highest \\
      & b-discriminant \\
 \drLB & Distance between the lepton and the \\
       & jet with the highest b-discriminant \\
\hline
\hline
\end{tabular}
\caption{List of discriminating variables used as input to a BDT in~\cite{st4bd}.}
\label{tab:vars}
\end{center}
\end{table}

\section{Quantum annealing and zooming}
\label{s:qamlz}

From the distribution of each variable $i$ in signal and background events,
we construct a weak classifier $\chi_i$ as in~\cite{nature} which retains
the discriminant character of each variable while adapting it to an annealing
process. We then construct an Ising problem as follows. For each training event
$\tau \in [1,S]$, we consider the vector $x_{\tau}$ of the values of each variable
of index $i$ we use, and a binary tag $y_{\tau}$ labeling the event $\tau$ as
either signal ($+1$) or background ($-1$). The value of the $i$th weak classifier
for the event $\tau$ is given by the sign of the corresponding weak classifier
$\chi_i$: $c_i(x_{\tau})=\mathrm{sgn}(\chi_i(x_{\tau}))/N=\pm1/N$, where $N$ is the number
of weak classifiers. In the QAML algorithm, the optimization of the signal-background 
classification problem is expressed in terms of the search for the set of spins
$s_i$ minimizing the Ising Hamiltonian:
\begin{eqnarray}
  H_{\mathrm{Ising}} & = & \sum\limits_{i=1}^{N} h_i s_i
  + \sum\limits_{i=1}^{N}\sum\limits_{j>i}^{N} J_{ij} s_i s_j \nonumber \\
  & = & \sum\limits_{i=1}^{N} \Big(\lambda - C_i + \frac{1}{2}\sum\limits_{j>i}^{N} C_{ij}\Big) s_i
  + \frac{1}{4}\sum\limits_{i=1}^{N}\sum\limits_{j>i}^{N} C_{ij} s_i s_j 
\label{eq:His}
\end{eqnarray}
where $h_i$ is the local field on spin $s_i$, and $J_{ij}$ is the coupling 
between spins $s_i$ and $s_j$. The factor $\lambda$ is a regularization constant, and 
the terms $C_i$ and $C_{ij}$ are defined as function of weak classifier 
values and event tags as:
\begin{eqnarray}
  C_{i} = \sum\limits_{\tau=1}^{S}c_i(x_{\tau})y_{\tau}, &
  C_{ij} = \sum\limits_{\tau=1}^{S}c_i(x_{\tau})c_j(x_{\tau}).
\label{eq:ci}
\end{eqnarray}
A strong classifier $R$ is then built as a linear combination of all weak 
classifiers and the spins, merging for each event the discriminating power 
provided by all $c_i$'s and the spins $s_i$ obtained from the quantum 
annealing process. The minimization of the classification error is 
performed by the minimization of the Euclidean distance between the binary 
tag of each event and its classification $R$ as obtained by the annealing:
\begin{eqnarray}
  || y - R ||^2 & = & \sum\limits_{\tau=1}^{S} |y_{\tau} - \sum\limits_{i=1}^{N} s_i c_i(x_{\tau})|^2.
\label{eq:norm}
\end{eqnarray}
In the QAML-Z approach, the quantum annealing is operated iteratively 
while a substitution is made to the spin $s_i$:
\begin{eqnarray}
s_i & \longrightarrow & \mu_i(t) + s_i \cdot \sigma(t) = \mu_i(t+1),
\label{eq:subs}
\end{eqnarray}
where:
\begin{itemize}
\item $\mu_i(t)$ is the mean value of qubit $i$ at time $t$. We have: 
$\forall i$ $\mu_i(0)=0$.
\item $\sigma(t)$ is the search width at each annealing iteration $t$. We 
have: $\sigma(t)=b^t$ where $b=\frac{1}{2}$ and $t \in [0,T-1]$.
\end{itemize}
This iterative procedure effectively shifts and narrows the region of 
search in the space of spins. It updates the vector $\mu_i$ which is 
collected at the final iteration to form the strong classifier:
\begin{eqnarray}
  R(x_{\tau}) & = & \sum\limits_{i=1}^{N} \mu_i(T - 1) c_i(x_{\tau}),
\label{eq:R}
\end{eqnarray}
where the use of the weak classifiers is not limited to the binary choice 
$\{0,1\}$, but is extended to the continuous interval $[-1,1]$ via the use 
of the vector $\mu_i$. The classification capacity of the QAML-Z algorithm is 
further enhanced by an augmentation scheme applied on the weak 
classifiers. For each $h_i$, several new classifiers $c_{il}$ are created:
\begin{eqnarray}
  c_{il}(x_{\tau}) & = & \frac{\mathrm{sgn}(h_i(x_{\tau}) + \delta l)}{N} ,
\label{eq:aug}
\end{eqnarray}
where $l \in \mathbb{Z}$ is the offset: $-A \leq l \leq A$, and $\delta$ 
is the step size. While the value $c_i$ of the old classifier has only a 
binary outcome for each $h_i$, the new classifiers $c_{il}$ have similar 
but $(2A+1)$ different outcomes depending on the very distribution of 
$h_i$. We therefore have a better discrimination because a more 
continuous, thus more precise representation of the spectrum of $h_i$ with 
$c_{il}$ than with $c_i$. Applying the substitution of Eq.~\ref{eq:subs} 
in Eq.~\ref{eq:norm}, omitting spin independent and quadratic self-spin 
interaction terms, and defining new indices $I$ as $\{il\}$ and $J$ as 
$\{jl'\}$, we obtain the Hamiltonian (see Appendix~\ref{s:hder}):
\begin{eqnarray}
  H(t) = \sum_{I=1}^{N(2A+1)}\left( -C_I + \sum_{J=1}^{N(2A+1)}\mu_J(t) C_{IJ}\right)\sigma(t) s_I
  + \frac{1}{2}\sum_{I=1}^{N(2A+1)}\sum_{J\neq I}^{N(2A+1)} C_{IJ}\sigma^2(t)s_I s_J, 
\label{eq:Hta}
\end{eqnarray}
with:
\begin{eqnarray}
  C_{I} = \sum\limits_{\tau=1}^{S}c_{il}(x_{\tau})y_{\tau}, &
  C_{IJ} = \sum\limits_{\tau=1}^{S}c_{il}(x_{\tau})c_{jl'}(x_{\tau}).
\label{eq:cia}
\end{eqnarray}
The terms $C_I$ and $C_{IJ}$ are the input to the classification problem. 
The Hamiltonian $H(t)$ is iteratively optimized for $t$, with the vector 
$\mu_I$ updated similarly to Eq.~\ref{eq:subs}. The information about the 
iterative quantum annealing, corresponding parameters, and control results 
are provided in Appendix \ref{s:qas}, where we ensure that the Ising model 
energy decreases and stabilizes for the chosen parameters. The output of 
the optimization procedure is a strong classifier built as in 
Eq.~\ref{eq:R}, and whose distribution is used to discriminate signal from 
background.

\section{Classification of stop with the QAML-Z algorithm}
\label{s:qastop}

As in \cite{st4bd}, only the main background processes \wjets and \ttbar 
are used for training the QAML-Z algorithm. To realistically represent the 
SM in the training, a background sample is formed where events of these 
two processes are present proportionally to their production rate at the 
LHC. We divide this sample in two equal parts, one being used by the 
QAML-Z algorithm and one to assess the performance of the strong 
classifier through the maximization of the FOM: 
N(Sample)=N($QA$)+N(Assess). The $QA$ sample is further divided in two 
equal parts, one to train the annealer and another one to test for 
over-training in the annealer: N($QA$)=N(Train)+N(Test). It should be 
noted that only the Train sample is involved in the annealing process. 
Having shown \cite{st4bd} that the kinematic properties of all signal 
points $(m(\tilde{t}_{1}),m(\tilde{\chi}_{1}^{0}))$ are quasi identical 
along the line $\Delta m = m(\tilde{t}_{1}) - m(\tilde{\chi}_{1}^{0})$, we 
use all signal events with $\Delta m = 30$ except the signal point 
$(550,520)$ as $QA$ sample, while entirely using this latter signal as 
Assess sample. This organization of samples allows the usage of a maximal 
number of both signal and background events for assessing the performance 
of the classification as well as testing the annealing process.

The data is run on the $2000Q$ quantum annealer of D-Wave Systems Inc. 
\cite{dwave}, where the time to solution is $O(\mu s)$, 
\emph{ie.} the time of the annealing (see Appendix \ref{s:qas}). This 
computer is based on the Chimera graph which has 2048 qubits and 5600 
couplers. To embed the Ising Hamiltonian in the annealer, qubits of the 
graph are ferromagnetically coupled into a chain to represent a single 
spin of the Hamiltonian $H(t)$. While the Hamiltonian in Eq.~\ref{eq:Hta} 
is fully connected, the Chimera graph is not, thus limiting the hardware 
implementation of the classification problem. The number of $J_{ij}$ 
couplers is given by: $N(J_{ij})=N_{\mathrm{v}}\cdot(N_{\mathrm{v}}-1)/2$ 
with $N_{\mathrm{v}}=N_{\mathrm{var}}\cdot(2A+1)$, where 
$N_{\mathrm{var}}$ is the number of variables used to train the QAML-Z 
algorithm, and $A$ is a parameter of the augmentation scheme (see 
Eq.~\ref{eq:aug}). Given the number of variables and the augmentation 
schemes used, the limit of 5600 couplers can be exceeded by $N(J_{ij})$; 
typically, for $N_{\mathrm{v}}$=12 and for an augmentation with $A$=5, the 
needed number of couplers is 8646. We therefore prune the elements of the 
$J_{ij}$ matrix, retaining the largest $(1-C)$ elements, where $C$ is a 
cutoff percentage. Different cutoff values are expected to optimize the 
performance for different sets of variables and different augmentations 
schemes ($A, \delta$). As a further option to reduce the size of the Ising 
model to be encoded on the annealer, we use the polynomial-time variable 
fixing scheme of the D-Wave API. This scheme is a classical procedure to 
fix the value of a portion of the input variables to values that have a 
high probability of being optimal. An illustration of the effect of the 
cutoff $C$, the use of variable fixing, and the augmentation scheme is 
given in Appendix \ref{s:qas}.

\begin{figure}[!htbp]
\begin{center}
\includegraphics[scale=0.28]{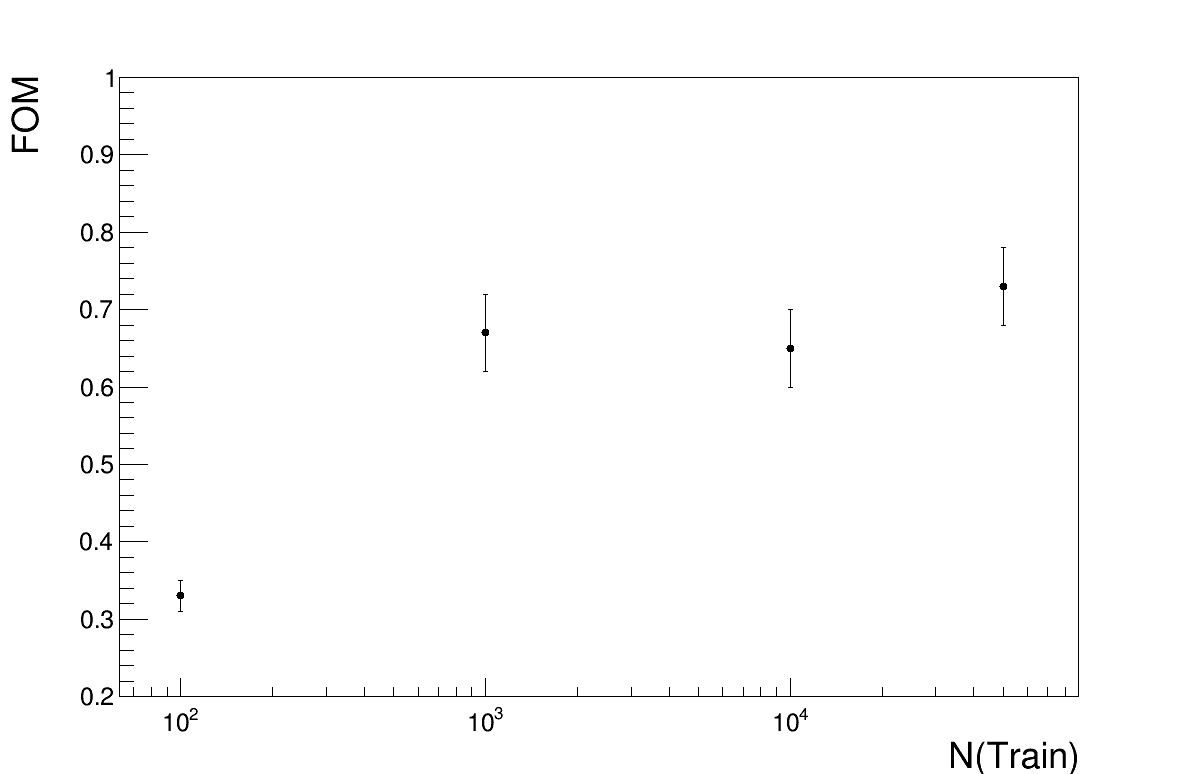}
\end{center} 
\caption{Evolution of the FOM as a function of the number events used for
  training. The QAML-Z algorithm uses the variables of Table~\ref{tab:vars} transformed
  in weak classifiers, with an augmentation scheme of ($\delta$,$A$)=(0.025,3),
  and with a cutoff $C$=85\%, without using a variable-fixing procedure.}
\label{fig:fom15}
\end{figure}

\begin{figure}[!htbp]
\begin{center}
\includegraphics[scale=0.5]{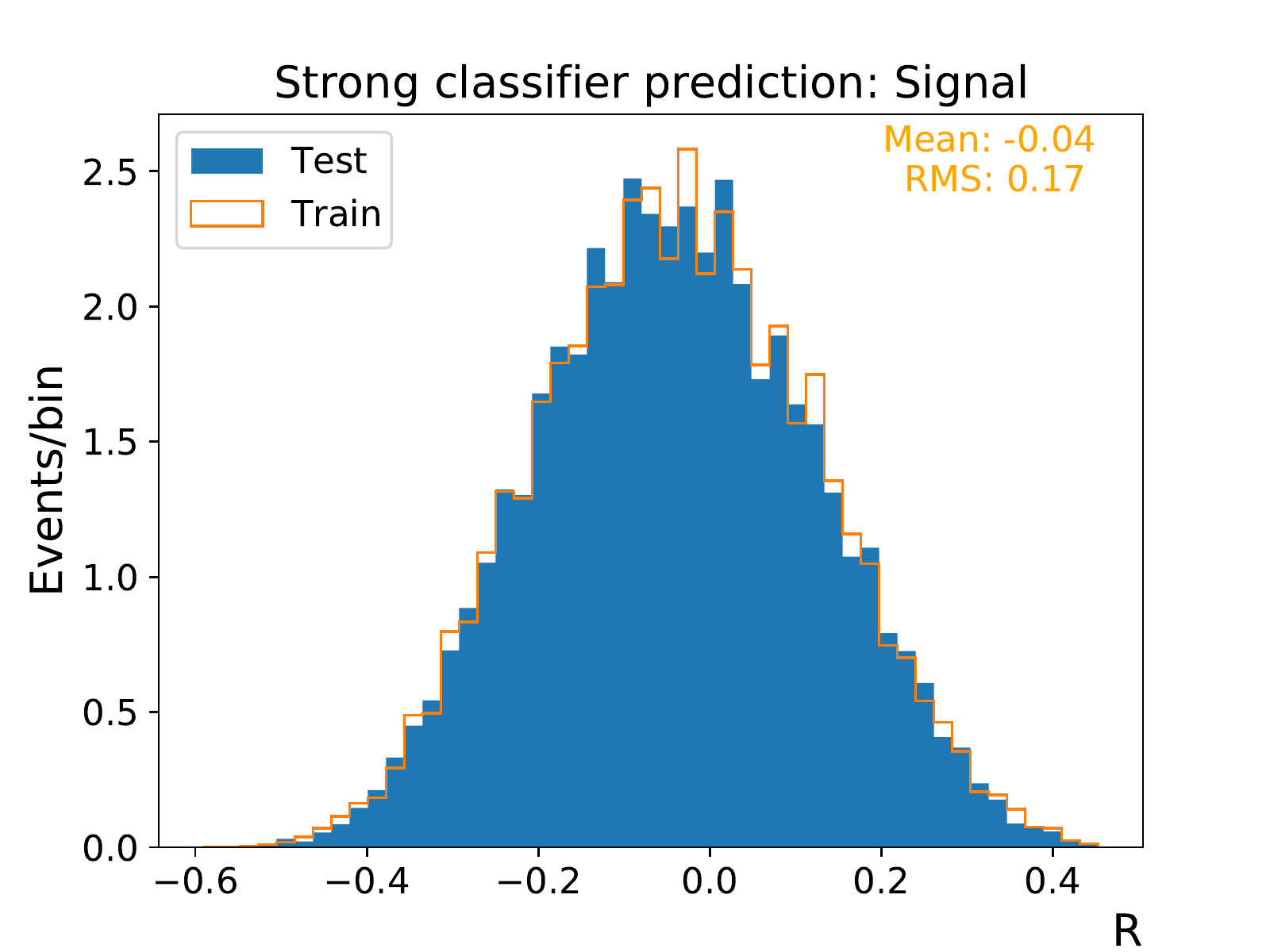}
\includegraphics[scale=0.5]{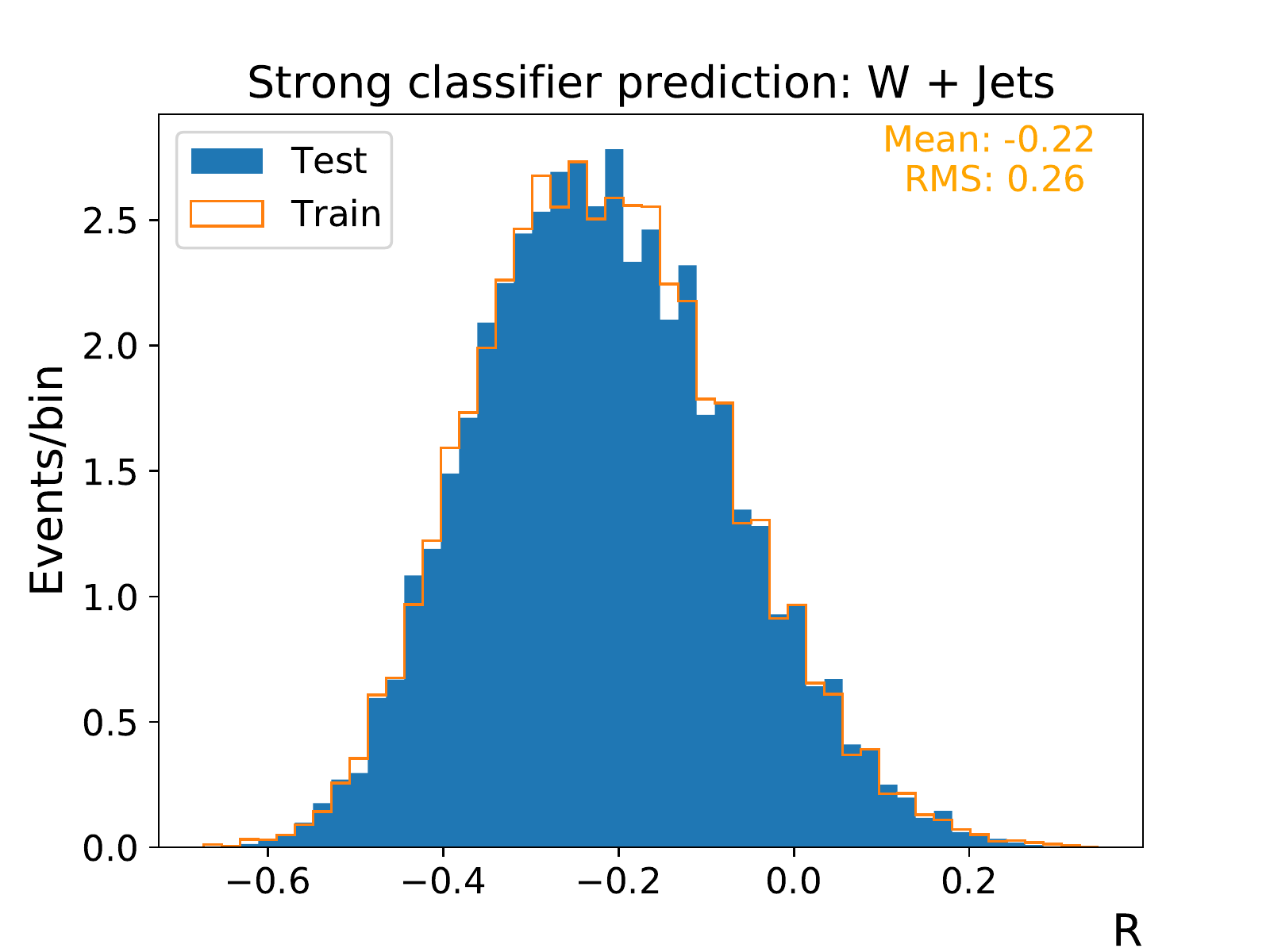} \\
\includegraphics[scale=0.5]{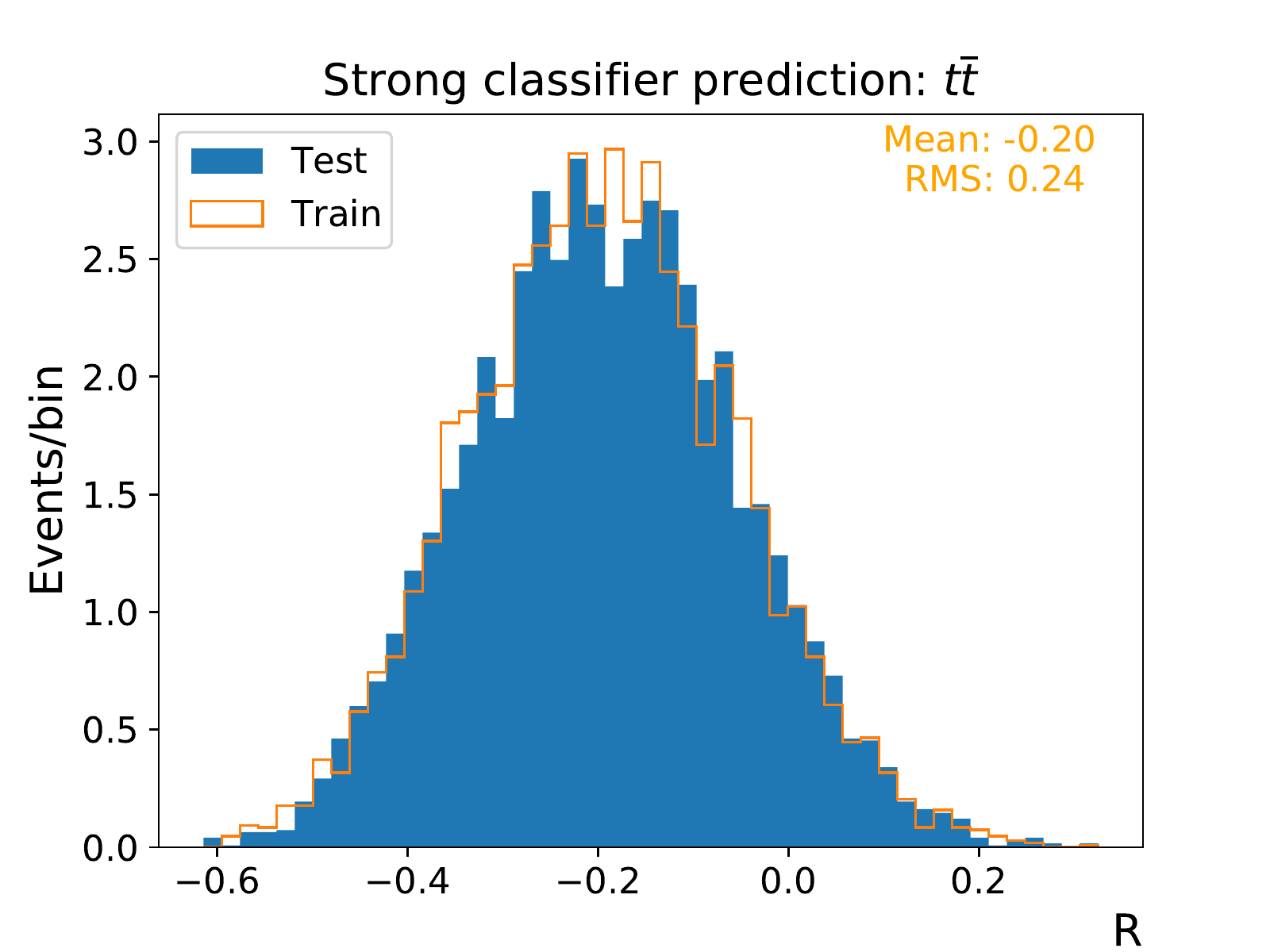}
\includegraphics[scale=0.5]{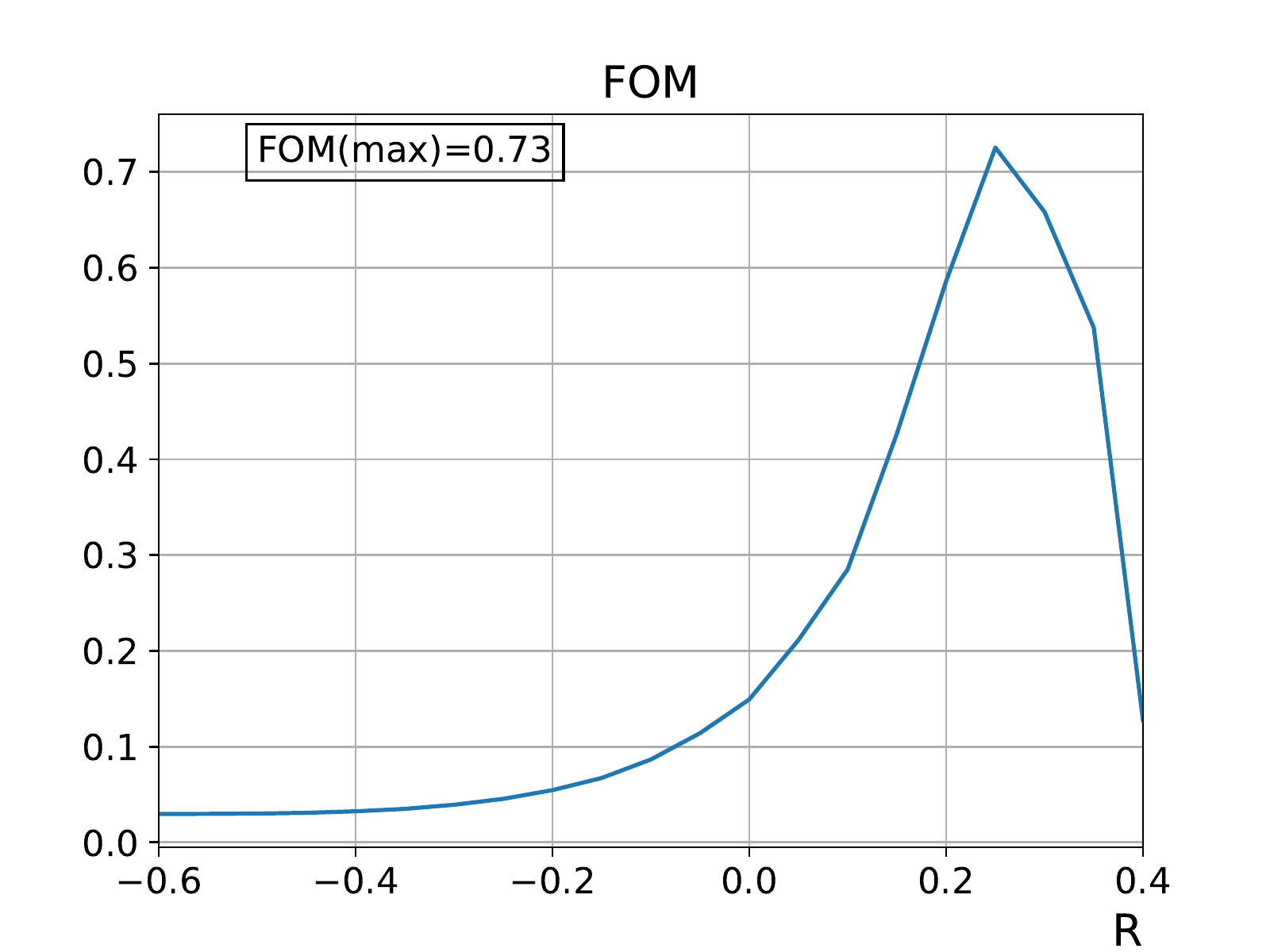}
\end{center} 
\caption{The output of the strong classifier for the signal (top-left),
  \wjets (top-right) and \ttbar background (bottom-left) in the train
  (orange) and test (blue) events within the $QA$ sample. The evolution
  of the FOM as a function of the cut applied on the strong classifier's
  output is illustrated in the plot in the bottom-right. The QAML-Z algorithm uses
  the variables of Table~\ref{tab:vars} transformed in weak classifiers,
  with an augmentation scheme of ($\delta$,$A$)=(0.025,3), and with a
  cutoff $C$=85\%, without using a variable-fixing procedure. The number
  of events used for training is N(Train)=50$\cdot$10$^3$.}
\label{fig:plots15}
\end{figure}

In order to compare the performance of a quantum annealing with a 
classical ML counterpart, we explore various $settings$ of the QAML-Z 
algorithm, namely different augmentation schemes, cutoffs, and variable 
fixing options, reporting only the performance of the best setting for 
each tested set of variables (see Sec.~\ref{s:res}). Despite averaging 
out the random errors on the annealing and mitigating the possible effects 
of overfitting due to zooming (see Sec.~\ref{s:qamlz}), the outcome of 
the annealing (the vector $\mu_I$) can vary due to the probabilistic 
nature of these schemes and to the variations of the machine itself (e.g. 
low-frequency flux noise of the qubits), leading to an uncertainty on the 
performance. In order to estimate this uncertainty, we run the annealing 
ten times with the same input variables, in the very same setting, and on 
the same sample of events, and we consider the standard deviation of the 
corresponding maximal FOMs as uncertainty of the performance for a given 
set of variables and setting. In Fig.~\ref{fig:fom15} we report the 
performance of the QAML-Z algorithm with the variables of 
Table~\ref{tab:vars} as input and with a given augmentation scheme and 
cutoff as a function of the number of events used in the training. The 
performance of the annealer increases with N(Train), witnessing a clear 
rise for rather small number of events and a more moderate increase for 
larger numbers of events, confirming the results of~\cite{nature} with 
another signal. Henceforth, we will present all results for 
N(Train)=N(Test)=50$\cdot$10$^3$ where signal and background events 
respectively represent 40\% and 60\% of these two samples. We therefore 
benefit from a large sample size to train the QAML-Z algorithm, while 
observing a quasi identical evolution of the Hamiltonian energy for the 
Train and Test samples (see Fig.~\ref{fig:enwf15} in Appendix 
\ref{s:qas}). The Assess sample contains approximately 200$\cdot$10$^3$ 
background, and 7$\cdot$10$^3$ signal events. In Fig.~\ref{fig:plots15} we 
present the distribution of the strong classifier for signal, and the two 
main background processes. As can be observed, there is no over-training 
of the QAML-Z algorithm because the response of the strong classifier is 
statistically very similar for events which are used to train the annealer 
and those not exposed to the training. Also shown in 
Fig.~\ref{fig:plots15} is the evolution of the FOM in the Assess sample as 
function of the cut applied on the output of the strong classifier. 
Henceforth, all the reported values of maximal FOM are checked to 
correspond to a cut where there are enough events in both signal and 
background samples.

\section{Approaches and results}
\label{s:res}

We define the main sets of tested variables in Table~\ref{tab:varset}. For each
set and the different approaches to test it, we perform an extensive study of
the performance of the QAML-Z algorithm for different augmentation schemes,
cutoffs, and the use (or not) of variable fixing, as illustrated for the sets
A and B respectively in Fig.~\ref{fig:fomA} and~\ref{fig:fomB} of Appendix
\ref{s:qas}. For each set and approach, we report the optimal setting and the
corresponding performance in Table~\ref{tab:fomAll}.

The set $\alpha$ contains the variables defined in Table~\ref{tab:vars} 
where the discriminating variables are not transformed into weak 
classifiers, being only normalized to the [$-1$,$+1$] interval. The set 
$\beta$ consists of the same variables where these are transformed into 
weak classifiers. As can be seen in Table~\ref{tab:fomAll}, the 
performance of the set $\beta$ is expectedly higher than for the set 
$\alpha$, where the weak classifiers are scaled as a function of the 
initial distribution of the discriminating variables to better reflect the 
separation between signal and background. The transformation into weak 
classifiers is performed for all subsequent tests.

We explore in a second step the effect of additional discriminating 
variables built from the same initial set of Table~\ref{tab:vars}. The 
methodology followed to built these new variables in explained in Appendix 
\ref{s:advar}, where the discriminating power of each variable is 
appraised via its maximal FOM. Two new sets of variables are constructed 
based on these new variables, as reported in Table~\ref{tab:varset}: the 
set A including the variables of the Table~\ref{tab:vars} and new 
variables with the highest FOMs, and the set B including those of set A 
and additional variables with the lower FOMs (see Table~\ref{tab:varnew}). 
As can be observed in Table~\ref{tab:fomAll}, the addition of variables 
with higher maximal FOMs in the set A increases the performance of the 
QAML-Z algorithm, while the further addition of variables with lower FOM in
the set B does not significantly improve the quality of the classification.

The results of the search~\cite{st4bd} are based on the use of BDT where 
the discriminating variables are diagonalized before being fed to the 
training~\cite{RefBDT,tmva}. This step better prepares the data for 
classification because the original discriminating variables do not 
necessarily constitute the optimal basis in which signal and background 
are optimally separated. In order to render our approach as comparable as 
possible to the one followed with a BDT~\cite{st4bd}, we pass our data 
through the procedure of PCA~\cite{pca} before feeding it to the QAML-Z
algorithm. It must be noted that the use of PCA is only one method for
diagonalizing the data, other methods also being applicable to this end.
The application of PCA on the data before the quantum annealing further
improves the results for the set of variables A, and to a lesser extent
for B, as can be seen in Table~\ref{tab:fomAll}. We note a larger uncertainty
of the QAML-Z algorithm where the data is prepared with the PCA, this for
the same sets of variables. In the PCA basis, the weak classifiers are
more decorrelated from each other, rendering the corresponding weights
$\mu_I$ more independent from one another. When a $\mu_I$ fluctuates (e.g.
because of the state of the machine), the strong classifier $R$
(see Eq.~\ref{eq:R}) is sensitive to the variations of a larger number of
$\mu_I$'s, hence a larger variation of its outcome. It is noticeable that
the QAML-Z algorithm, once put on a footing as similar as possible to the
BDT based approach~\cite{st4bd}, can reach an equivalent, possibly better
performance. It is interesting to observe that the best result is achieved
without using the variable fixing scheme, where the annealing is put at
full use.

\begin{table}[!htbp]
\begin{center}  
\begin{tabular}{|l|l|l|}
\hline
\hline
Variable & List of   & Use of weak \\
set name & variables & classifiers \\
\hline
 $\alpha$ / $\beta$ & Table~\ref{tab:vars} & No / Yes  \\
 A  & Table~\ref{tab:vars} and:  & Yes  \\
    & \ptl/\met, \ptl/\ptisr, & \\
    & (\bdisc$-$1)\ptb,  & \\
    & $|$(\met$-$280)(\mt$-$80)$|$ , & \\
    & $|$(\met$-$280)(\Ht$-$400)$|$ & \\
 B  & Variables of set A and: & Yes \\
    & \drLB$-$ (\mt/40) , & \\
    & \Ht$^2$/\njet, \pt~$+$ 3.5$\eta(l)^2$ , & \\
    & \pt/\Ht & \\
\hline
\hline
\end{tabular}
\caption{Definition of different variable sets as a function of the used 
  variables.}
\label{tab:varset}
\end{center}
\end{table}

\begin{table}[!htbp]
\begin{center}  
\begin{tabular}{|l|c|c|c|c|c|}
\hline
\hline
Variable  & Fixing   & $C$ [\%] & ($\delta$,$A$) & FOM \\
  set     & variable &          &                &     \\
\hline
 $\alpha$ & False & 85 & (2.50$\cdot$10$^{-2}$,3) &  0.48 $\pm$ 0.03 \\
 $\beta$  & False & 85 & (2.50$\cdot$10$^{-2}$,3) &  0.73 $\pm$ 0.03 \\
 A        & True  & 95 & (0.90$\cdot$10$^{-2}$,5) &  0.88 $\pm$ 0.04 \\
 B        & True  & 85 & (0.70$\cdot$10$^{-2}$,3) &  0.91 $\pm$ 0.05 \\
 PCA(A)   & False & 95 & (1.45$\cdot$10$^{-2}$,3) &  1.57 $\pm$ 0.24 \\
 PCA(B)   & True  & 95 & (0.70$\cdot$10$^{-2}$,5) &  1.09 $\pm$ 0.17 \\
\hline
 BDT      & NA    & NA & NA        &  1.44 $\pm$ 0.06 \\
\hline
\hline
\end{tabular}
\caption{Best performance obtained for different sets of variables as defined
  in Table~\ref{tab:varset}, and for different approaches applied on some sets.
  The corresponding use (or not) of variable fixing, cutoff and augmentation
  scheme are reported. All results are provided for N(Train)=50000. For
  comparison, the performance of the BDT of~\cite{st4bd} is also reported,
  where ``NA'' stands for nonapplicable.}
\label{tab:fomAll}
\end{center}
\end{table}

\section{Summary}
\label{s:conc}

We studied the capability of the quantum annealing, where the zoomed and
augmented QAML-Z approach is applied to a new classification problem, namely
the discrimination of stop versus SM background events. The classification
is based on well motivated variables whose discriminating power has been
tested with a FOM maximization procedure. The use of this latter metric
constitutes a novel and reliable assessment of the performance of a selection
as it includes its full statistical and systematic uncertainties. We
systematically tested each set of variables used by the QAML-Z algorithm
for different augmentation schemes and percentages of pruning on
the couplers of the annealer as to find the optimal setting. The performance
of different settings is assessed for large training samples which are
observed to yield the best performances, and are also more adapted to the
needs of experimental particle physics where very large data samples are
used. We observe an improvement of the classification performance when
adding variables with a high FOM. To put the annealing approach and the
classical BDT approach on the same footing, we pass the data through a PCA
procedure before feeding it to the quantum annealer. For the first time in
HEP, we show that for large training samples the QAML-Z approach running on
the Chimera graph reaches a performance which is at least comparable to the
best-known classical ML tool. With more recent graphs there is the prospect that the
larger number of connected qubits will yield a better correspondence between
the Ising Hamiltonian and the system of qubits of the annealer. The larger
number of available couplers in the machine will allow a more complete use
of the information contained in the couplers of the Hamiltonian; it will
render each chain more stable, thus less prone to be broken, where the
discriminating information of the classification will be more effectively used.

\section*{Acknowledgements}

We acknowledge the authors of \cite{nature,qamlz} for their help which has 
been key for this study. We specially thank the Center for Quantum 
Information Science \& Technology of the University of Southern California 
in Los Angeles for granting us access to the $2000Q$ quantum annealer of 
D-Wave Systems Inc. \cite{dwave}. P.B. thanks the support from 
Funda\c{c}\~{a}o para a Ci\^{e}ncia e a Tecnologia (Portugal), namely 
through project UIDB/50008/2020, as well as from project $QuantHEP$
$Quantum$ $Computing$ $Solutions$ $for$ $High$-$Energy$ $Physics$, 
supported by the EU H2020 QuantERA ERANET Cofund in Quantum Technologies, 
and by FCT (Grant No. QuantERA/0001/2019).

\bibliographystyle{plain}


\appendix

\section{Samples and signal selection}
\label{s:pb16}

The data used for training and testing the QAML-Z algorithm are events 
simulating proton-proton collisions of the LHC at $\sqrt{s}$ = 13 TeV. The 
generation of the signal and background processes is performed with the 
\textsc{Madgraph5}2.3.3~\cite{MG} generator. All samples are then passed 
to \textsc{Pythia}~8.212~\cite{Pythia1,Pythia2} for hadronization and 
showering. The detector response is simulated with the 
\textsc{Delphes}.3~\cite{DE} framework.

The first step in the search is the preselection, which is common to the 
search~\cite{st4bd} and to the present study. We require \met$>280$ GeV to 
select preferentially signal, as the production of two \lsp escaping 
detection increases the missing transverse energy in the detector. The 
leading jet of each event is required to fulfill \pt$>110$ GeV and 
$|\eta|<2.4$. A cut of \Ht$>200$ GeV is imposed. This cut diminishes the 
contribution of the \wjets background where jets are softer than for 
signal. At least one identified muon (electron) with \pt $>3.5$ $(5)$ GeV 
and $|\eta|<2.4$ ($2.5$) must be present. Events with additional leptons 
with \pt $>$ 20 GeV are rejected, diminishing the contribution of the 
\ttbar background with two leptons. Background from SM dijet and multijet 
production are suppressed by requiring the azimuthal angle between the 
momentum vectors of the two leading jets to be smaller than 2.5 rad for 
all events with a second hard jet of \pt$>60$ GeV.

The second step in the selection of the signal is the usage of an approach 
more advanced than linear cuts. We describe here the parameters of the BDT 
as used in~\cite{st4bd}, and the procedure to include different 
discriminating variables as its inputs. We define $N_T$ and $M_D$ 
respectively as the number of trees in the BDT and its maximal depth. The 
maximal node size $M_N$ is the percentage of the number of signal or 
background events at which the splitting of data stops in the tree, and 
acts as a stopping condition of the training. These parameters are 
optimized by maximizing the FOM of different BDTs trained with various 
parameters. The setting yielding the best performance while avoiding an 
over-training is ($N_T$, $M_D$, $M_N$)=(400, 3, 2.5\%). Finally, the space 
of input variables is diagonalized before being fed to the training. As 
for the procedure to include discriminating variables as inputs to the 
BDT, we start from a reduced set $\xi$ which comprises the basic variables 
of the search. A new variable $v$ is incorporated into the set of input 
variables only if it significantly increases the FOM. Namely, we train a 
BDT with the set $\xi$ and another with $\xi \oplus v$, and calculate the 
FOM of Eq.~\ref{eq:fom} as a function of the cut applied on the BDT's 
output. If the maximal FOM reached with the latter set is higher than the 
one with the former, the variable $v$ is incorporated as a new input 
variable; if the performance of the set $\xi \oplus v$ is compatible with 
the one of $\xi$, it is not. This procedure is repeated until there is no 
new variable at disposal.

In order to make the comparison with the results of quantum annealing as 
valid as possible, a BDT is re-trained with the \textsc{Delphes} 
simulation. The performance, for the same signal, is compatible between 
this new simulation and the full simulation of the CMS detector.

\section{Derivation of the Hamiltonian}
\label{s:hder}

In this section, we derive the expression of the Hamiltonian to be 
iteratively minimized, first for the zooming and then for the augmentation 
step. If we expand the Euclidean distance of Eq.~\ref{eq:norm} to be 
minimized, we obtain:
\begin{equation}
\sum_{\tau=1}^S\left[ ||y_{\tau}||^2 + \left(\sum_{i=1}^N s_i c_i\left(\textbf{x}_{\tau}\right)\right)^2 - 2y_{\tau}\left(\sum_{i=1}^N s_i c_i\left(\textbf{x}_{\tau}\right)\right)\right].
\label{eq:s-exp}
\end{equation}
Omitting the first term which is constant, and inserting the zooming 
substitution of Eq.~\ref{eq:subs} in Eq.~\ref{eq:s-exp}, we obtain:
\begin{equation}
\label{eq:s-exp-z}
\sum_{\tau=1}^S\left[\left(\sum_{i=1}^N (\sigma(t)s_i + \mu_i(t)) c_i(\textbf{x}_{\tau})\right)^2 -
  2y_{\tau}\left(\sum_{i=1}^N (\sigma(t)s_i + \mu_i(t)) c_i(\textbf{x}_{\tau})\right)\right].
\end{equation}
Fully developing the Eq.~\ref{eq:s-exp-z} while neglecting constant, spin 
independent, and quadratic self-spin interaction terms, we get:
\begin{equation}
\sum_{\tau=1}^S\left[
  - 2y_{\tau}\sum_{i=1}^N \sigma(t)c_i(\textbf{x}_{\tau})s_i
  + 2\sum_{i=1}^{N} \left( \sum_{j=1}^N \mu_j(t) c_j(\textbf{x}_{\tau}) \right) \sigma(t) c_i(\textbf{x}_{\tau}) s_i
  + 2\sum_{i=1}^{N}\sum_{j>i}^{N} (\sigma^2(t) c_i(\textbf{x}_{\tau})c_j(\textbf{x}_{\tau}))s_is_j
  \right].
\label{eq:spin-dep-terms}
\end{equation}
Recalling the definition of the terms $C_i$ and $C_{ij}$ in 
Eq.~\ref{eq:ci}, and dividing the expression~\ref{eq:spin-dep-terms} by 
two, we obtain the expression of the Hamiltonian with the zooming 
approach:
\begin{eqnarray}
  H(t) & = & 
  \sum\limits_{i=1}^{N} \Big(- C_{i} + \sum\limits_{j=1}^{N}\mu_{j}(t)C_{ij}\Big) \sigma(t) s_{i}
  + \sum\limits_{i=1}^{N}\sum\limits_{j>i}^{N}C_{ij} \sigma^2(t) s_{i} s_{j}.
\label{eq:Htz}
\end{eqnarray}

Now we augment each classifier $i$ with $(2A + 1)$ classifiers:
\begin{eqnarray}
\forall i , \exists l & / & c_i \longrightarrow c_{il} , 
\label{eq:subaug}
\end{eqnarray}
where $l \in \mathbb{Z}$ is the offset: $-A \leq l \leq A$, as defined in 
Eq.~\ref{eq:aug}. Inserting the augmentation substitution~\ref{eq:subaug} in 
Eq.~\ref{eq:Htz} and omitting constant terms, we obtain:
\begin{eqnarray}
  H(t) = \sum_{l=-A}^{+A}\sum_{i=1}^N\left( -\sum_{\tau=1}^{S} c_{il}y_{\tau} + \sum_{l'=-A}^{+A}\sum_{j=1}^N\mu_{jl'}(t) \sum_{\tau=1}^S c_{il}c_{jl'}\right)\sigma(t) s_{il} + \frac{1}{2}\sum_{l=-A}^{+A}\sum_{i=1}^N \sum_{\{j,l'\} \neq \{i,l\}} \sum_{\tau=1}^S c_{il}c_{jl'} \sigma^2(t)s_{il} s_{jl'}.
\label{eq:Htza0}
\end{eqnarray}
Using the equivalence:
\begin{eqnarray}
\forall X , & \sum\limits_{i=1}^{N} \sum\limits_{l=-A}^{A} X_{il} & \equiv \sum\limits_{I=1}^{N(2A+1)} X_I,
\label{eq:equiv}
\end{eqnarray}
and defining the new indices $I$ and $J$ as $\{il\}$ and $\{jl'\}$ 
respectively, we obtain the expression of the final Hamiltonian which 
includes the zooming and augmentation approaches:
\begin{equation}
  H(t) = \sum_{I=1}^{N(2A+1)}\left( -C_I + \sum_{J=1}^{N(2A+1)}\mu_J(t) C_{IJ}\right)\sigma(t) s_I +
  \frac{1}{2}\sum_{I=1}^{N(2A+1)}\sum_{J\neq I}^{N(2A+1)} C_{IJ}\sigma^2(t)s_I s_J, 
\label{eq:Htza}
\end{equation}
with:
\begin{eqnarray}
  C_{I} = \sum\limits_{\tau=1}^{S}c_{il}(x_{\tau})y_{\tau}, &
  C_{IJ} = \sum\limits_{\tau=1}^{S}c_{il}(x_{\tau})c_{jl'}(x_{\tau}). 
\label{eq:ciag}
\end{eqnarray}
It has to be noted that the Hamiltonians of both equations \ref{eq:Htz} and
\ref{eq:Htza} are fully connected.

\section{Quantum annealing: parameters and control results}
\label{s:qas}

During the iterative optimization of the Hamiltonian, and to average out 
random errors on the local fields and couplings, each annealing is run and 
averaged over $n_g$ gauges \cite{gauge} and $n_e$ maximal number of 
excited states, where $n_g$ and $n_e$ can be made to monotonically 
decrease with each iteration. To mitigate the impact of overfitting due 
to the zooming, we follow a two-step randomization procedure in each 
iteration: if the energy of the qubit $i$ worsens, a sign flip $s_i 
\rightarrow -s_i$ is applied with a monotonically decreasing probability 
$p_f(t)$, followed by a randomly uniform spin flip for all qubits with 
probability: $q_f(t) < p_f(t)$ $\forall t$; the values of the two 
probabilities per iteration are kept the same as in~\cite{qamlz}. Building 
on the results of~\cite{nature,qamlz}, we set the number of iterations to 
8, while setting the annealing time to 20 $\mu s$. During this iterative 
optimization, the annealing is run for $n_g$ gauges at each iteration to 
reduce random errors on the local fields $h_i$ and couplers $J_{ij}$. For 
each gauge, the annealing result is sampled 200 times, and the set of 
spins leading to the lowest energy is collected. The selection of the 
excited states is based on a distance $d$ to the state of lowest energy 
and a maximal number $n_e$ of excited states as in ~\cite{qamlz}, where 
the value of $n_g$, $n_e$ and $d$ varies with the iteration. This means 
that after the iteration $t$ we have a set of $n_g(t)$ different 
$\mu_I$'s, out of which at most $n_e(t)$ are kept for the next iteration, 
corresponding to the best energies. At iteration $t+1$, we have at most 
$n_e(t) \cdot n_g(t)$ annealings, thus states, out of which at most 
$n_e(t+1)$ are kept. This represents a lot of computing time. For the 
problem optimized in this paper, and for a given set of variables, we 
compared results obtained with $n_g=\{50,10,1,...,1\}$ and 
$n_e=\{16,4,1,...,1\}$ as in~\cite{qamlz} on one hand, with results 
obtained with $n_g=\{50,10,...,10\}$ and $n_e=\{1,...,1\}$ on the other. 
We observed that the obtained energies were quasi-identical for a large 
number of training events, showing that picking the state of best energy 
out of $n_g > 1$ is sufficient to mitigate the uncertainties of the 
annealing process, while saving computing time. We therefore retain the 
latter options for $n_g$ and $n_e$. Finally, the chain strength $r$ is 
defined as the ratio of the coupling within each chain over the largest 
coupling in Hamiltonian. If $r$ is very large, the chains will be too 
strong to allow a multiqubit flipping which is necessary to explore the 
space of spins. If on the contrary $r$ is very small, the chains will be 
broken by the tension induced by the problem or by thermal excitation. The 
chain strength can be set to decay monotonically with each iteration as to 
allow $H(t)$ to drive the system dynamics while preventing the chains of 
qubits from breaking \cite{cstrength}. The value of $r$ at each iteration 
is the same as in~\cite{qamlz}. In the case where the chain is broken, the 
measure of the qubit chains is performed through a majority vote, possibly 
leading to the collection of non-optimal sets of solution spins, thus to a 
possible loss of discriminating information.

\begin{figure}[!htbp]
\begin{center}
\includegraphics[scale=0.4]{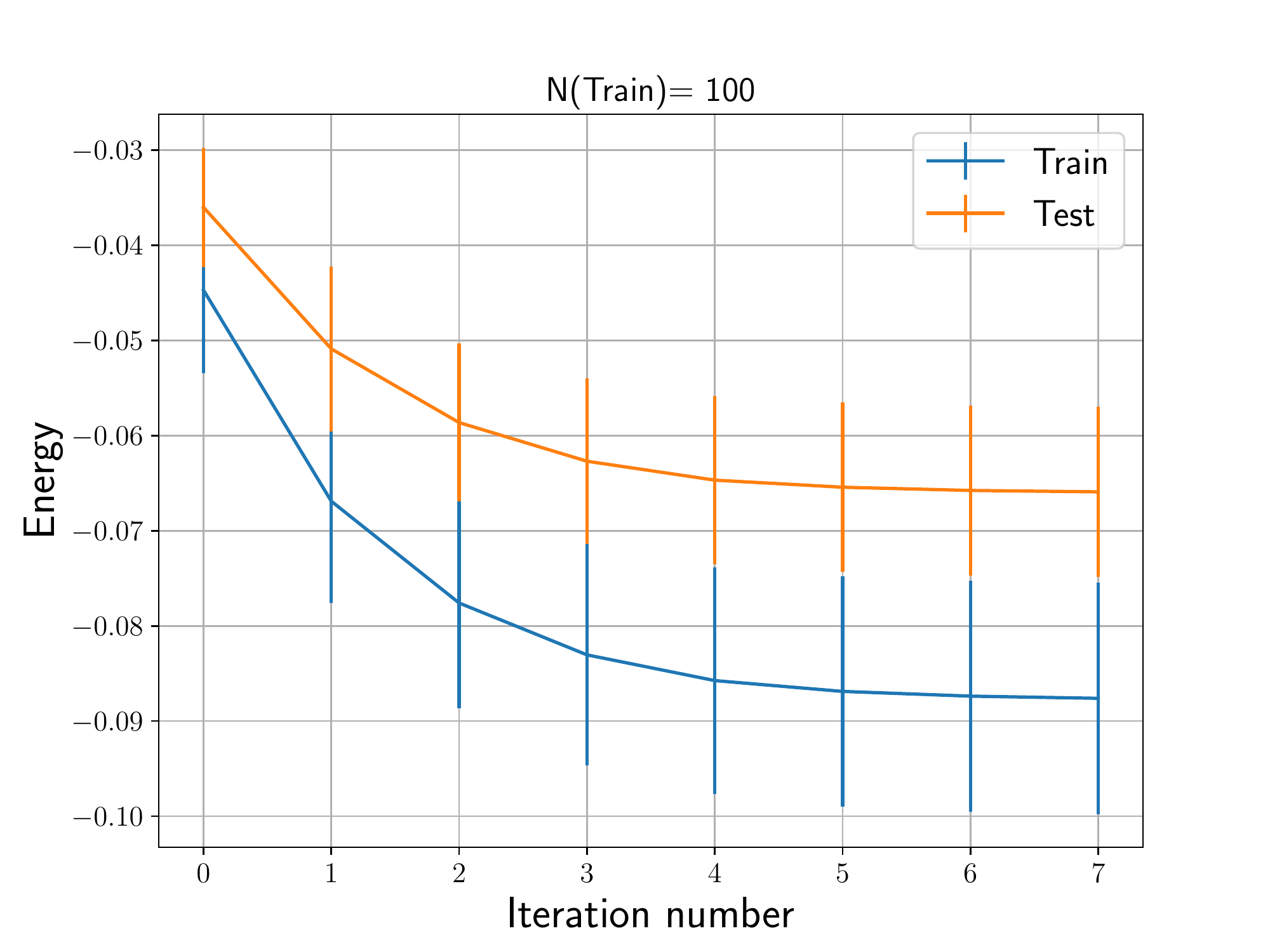}
\includegraphics[scale=0.4]{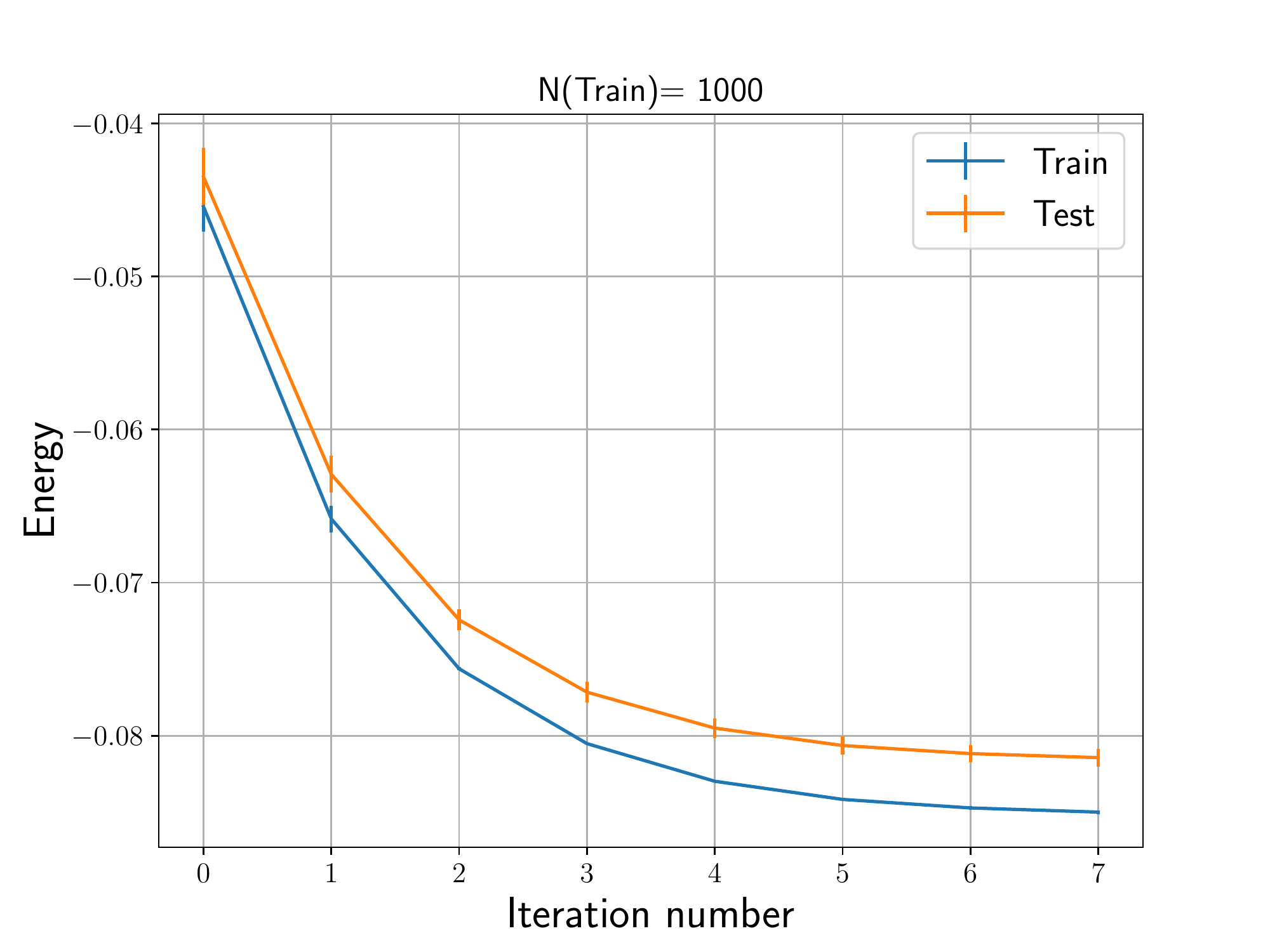} \\
\includegraphics[scale=0.4]{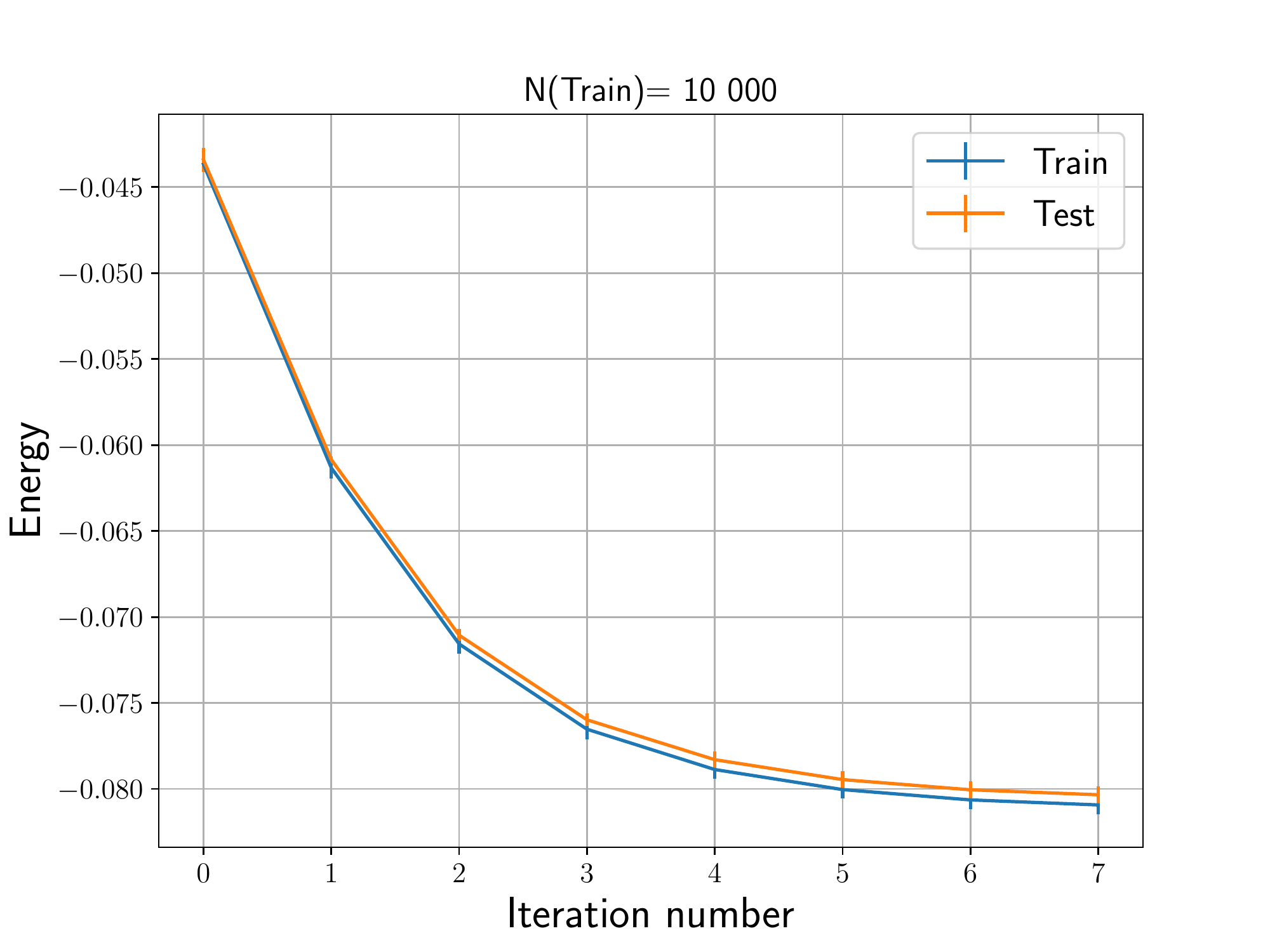}
\includegraphics[scale=0.4]{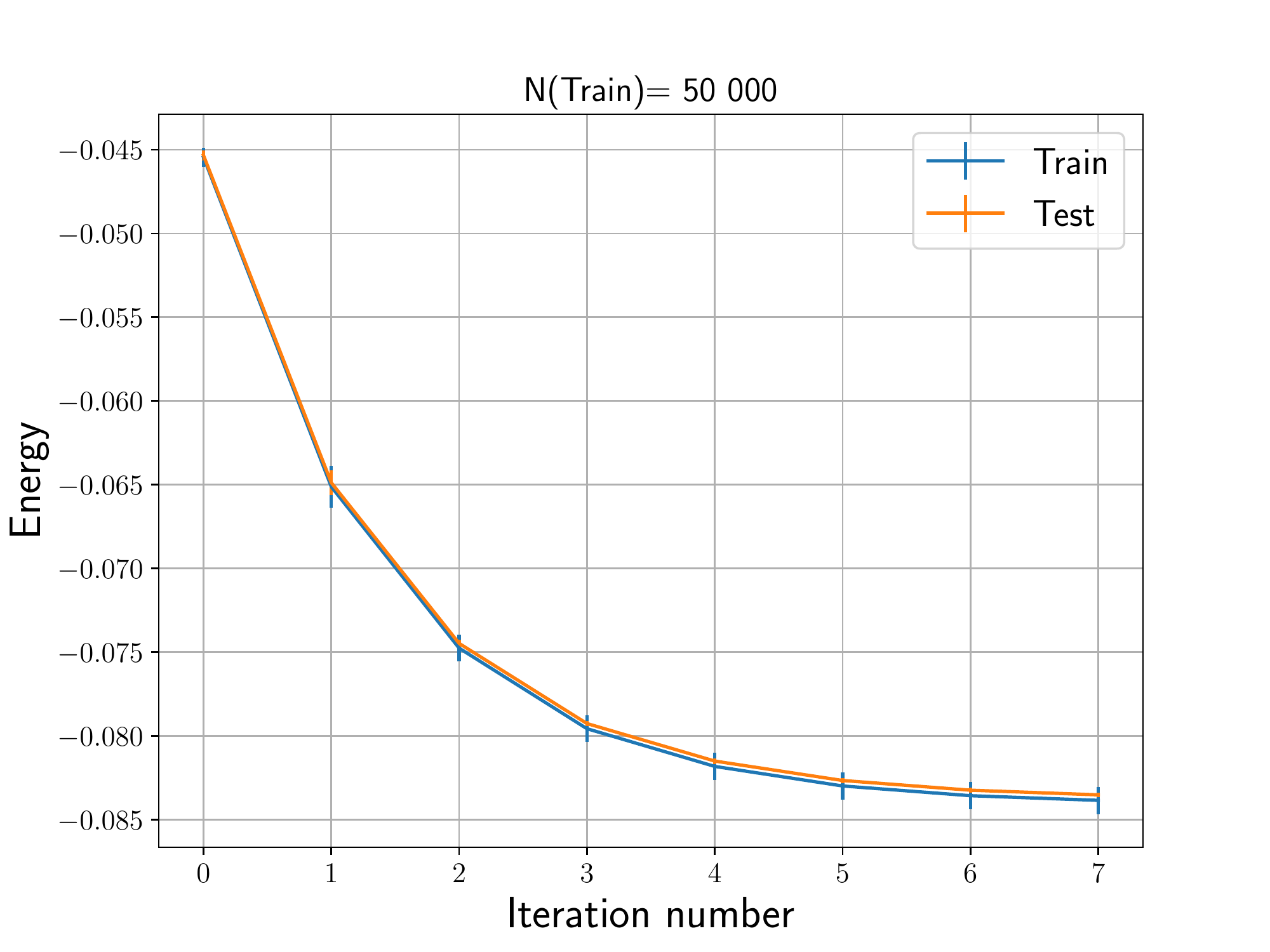}
\end{center}
\caption{Evolution of the Ising Hamiltonian energy as a function of the
  iteration for N(Train) varying from 100 (top left) to 50000 (bottom right),
  these in the train (blue) and test (orange) samples.
  The QAML-Z algorithm uses the variables of Table~\ref{tab:vars} transformed in weak
  classifiers, with an augmentation scheme of ($\delta$,$A$)=(0.025,3) and
  a cutoff of 85\%, without using a variable-fixing procedure.}
\label{fig:enwf15}
\end{figure}

\begin{figure}[!htbp]
\begin{center}
\includegraphics[scale=0.19]{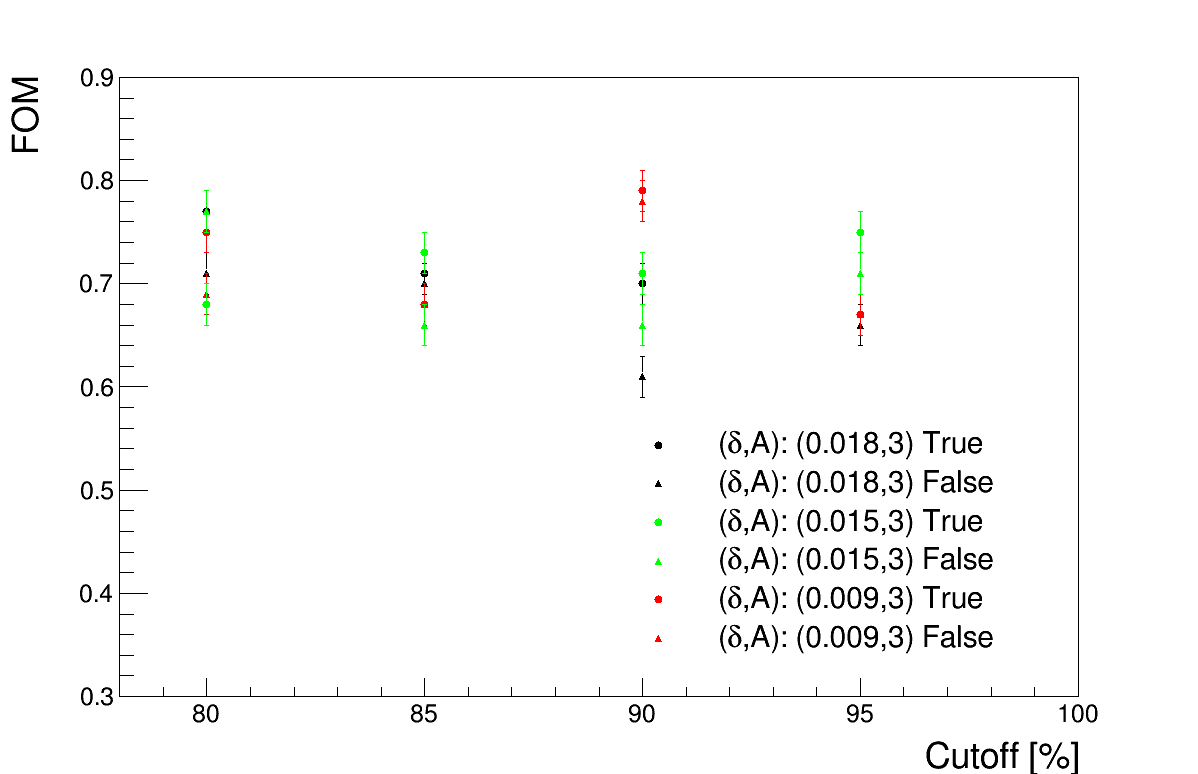}
\includegraphics[scale=0.19]{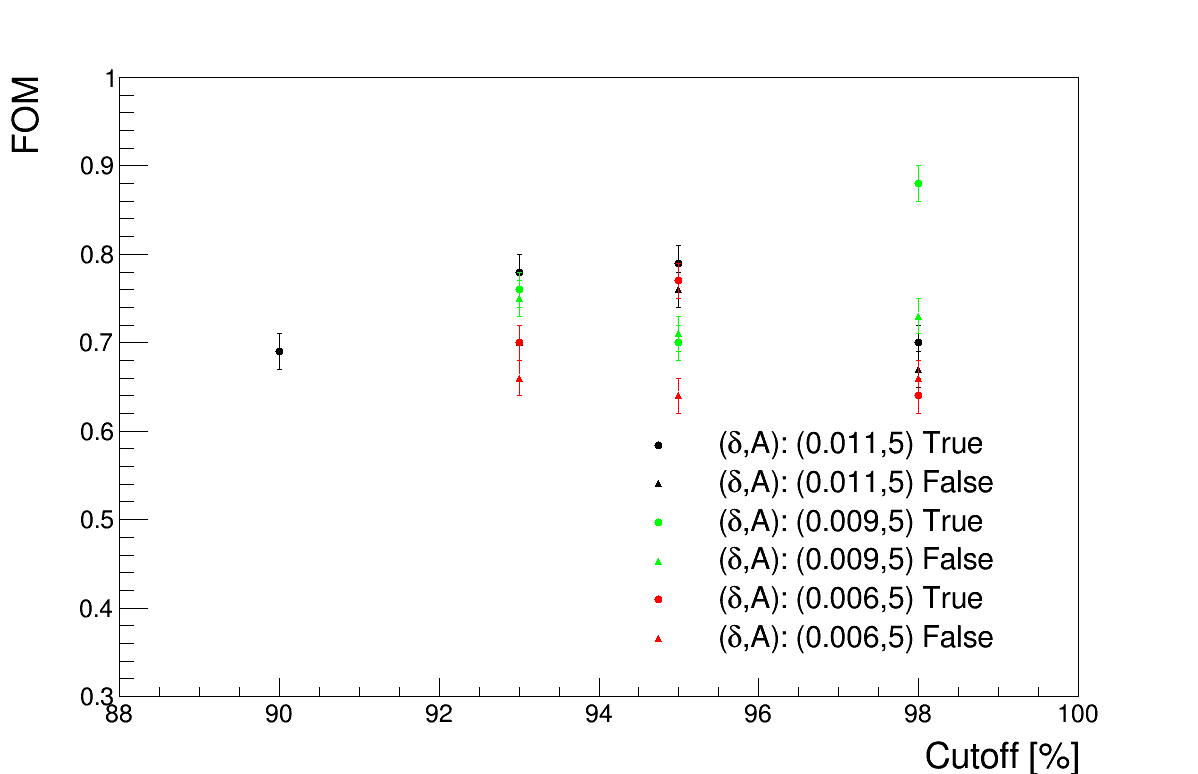}
\end{center} 
\caption{Evolution of the FOM as a function of the cutoff $C$ and for variable-fixing
  set to true (full circle) and false (full triangle). The QAML-Z algorithm uses the variable set A
  of Table~\ref{tab:varset}. The left plot presents results for an augmentation scheme of
  ($\delta$,$A$)=(0.018,3) in black, (0.015,3) in green and (0.009,3) in red. The
  right plot shows results for ($\delta$,$A$)=(0.011,5) in black, (0.009,5) in green
  and (0.006,5) in red.}
\label{fig:fomA}
\end{figure}

\begin{figure}[!htbp]
\begin{center}
\includegraphics[scale=0.19]{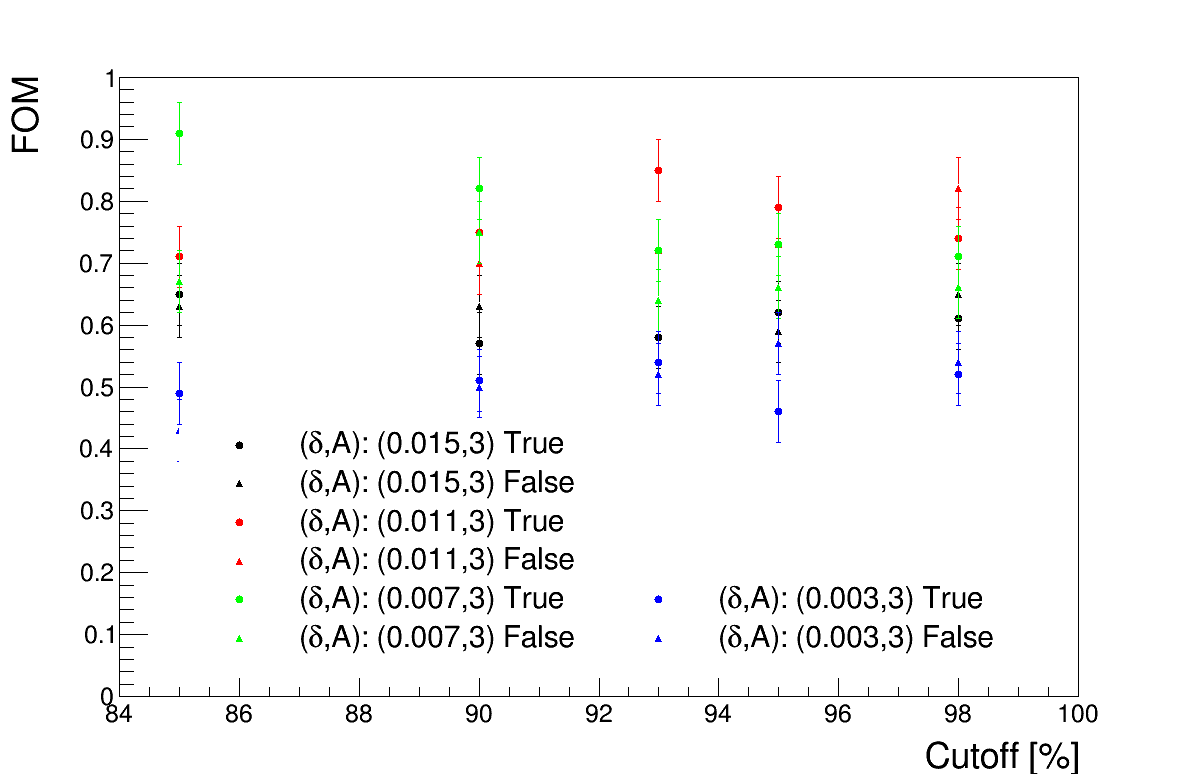}
\includegraphics[scale=0.19]{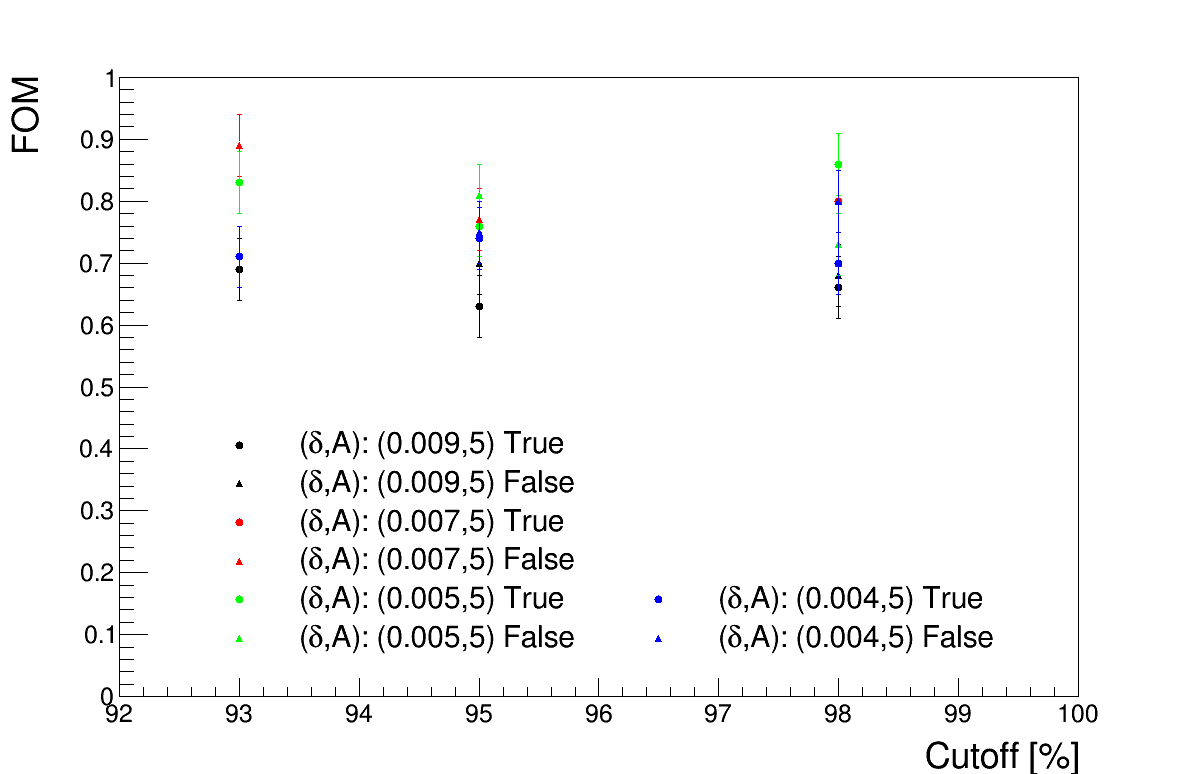}
\end{center} 
\caption{Evolution of the FOM as a function of the cutoff $C$ and for variable-fixing
  set to true (full circle) and false (full triangle). The QAML-Z algorithm uses the variable set B
  of Table~\ref{tab:varset}. The left plot presents results for an augmentation scheme of
  ($\delta$,$A$)=(0.015,3) in black, (0.011,3) in red, (0.007,3) in green, and
  (0.003,3) in blue. The right plot shows results for ($\delta$,$A$)=(0.009,5)
  in black, (0.007,5) in red, (0.005,5) in green, and (0.004,5) in blue.}
\label{fig:fomB}
\end{figure}

In Fig.~\ref{fig:enwf15} we present the evolution of the Ising 
Hamiltonian energy as a function of the optimization iteration for 
different numbers of events used to train the QAML-Z algorithm. One can observe that 
the energy decreases with the iteration, and that the difference of energy 
between events used to train and test the annealer decreases for higher 
N(Train). In figures \ref{fig:fomA} and \ref{fig:fomB} we report the 
performance of the QAML-Z algorithm for variable sets A and B as defined in 
Table~\ref{tab:varset}, and for different augmentation schemes, cutoffs 
$C$ and variable-fixing options. Lower values of $C$ than those 
illustrated in these figures are not reported because no embedding was 
found given the number of variables and tested augmentation scheme. It is 
interesting to note that the lowest bound for $C$ is higher for the set B 
where the number of variables is higher, and for a higher augmentation 
range (variable $A$) which leads to a larger number of $c_{il}$ (see 
Eq.~\ref{eq:aug}), thus coupling terms $C_{IJ}$ (see Eq.~\ref{eq:cia}). 
Depending on the set of variables and the augmentation scheme, different 
values of $C$ are optimal. The performance is generally higher when the 
variable-fixing procedure is used. For a given set of variables, small or 
big values of the offset $\delta$ (see Eq.~\ref{eq:aug}) might lead to a 
disadvantageous augmentation of the weak classifiers, leading to 
non-optimal performance. This is illustrated for the performances of the 
set B in Fig.~\ref{fig:fomB} where the FOM raises then drops for 
decreasing values of $\delta$, this for almost all values of cutoff.

\section{Advanced variables}
\label{s:advar}

We construct new variables by performing operations between the 
discriminating variables of Table~\ref{tab:vars}, where each new variable 
is built from two variables of this list. A two-dimensional distribution 
for each pair of initial variables allows to gauge the separation between 
signal and various background processes. Several analytic functions of the 
two variables are considered, and the one allowing to reach the highest 
FOM is retained per pair of variables. Seventeen new variables are 
constructed with this method, out of which nine are considered, see 
Table~\ref{tab:varnew}. Based on the discriminating power of each new 
variable, we consider two groups of variables, one made of the five new 
variables with the highest FOMs, and a second with the four subsequent 
ones.

\begin{table}[!htbp]
\begin{center}  
\begin{tabular}{|l|l|}
\hline
\hline
Variables & FOM \\
\hline
\ptl/\met & 0.35 \\
\ptl/\ptisr & 0.22 \\
(\bdisc$-$1)\ptb & 0.20 \\
$|$(\met$-$280)(\mt$-$80)$|$ & 0.20 \\
$|$(\met$-$280)(\Ht$-$400)$|$ & 0.18 \\
\hline
\drLB$-$ (\mt/40) & 0.12 \\
\Ht$^2$/\njet & 0.09 \\
\pt~$+$ 3.5$\eta(l)^2$ & 0.08 \\
\pt/\Ht & 0.03 \\
\hline
\hline
\end{tabular}
\caption{New discriminating variables constructed via operations on
  the ones of Table~\ref{tab:vars}. For each new variable, the maximal
  value of the FOM is reported, where the corresponding uncertainty is
  negligible. Variables of Table~\ref{tab:vars} and those with higher
  FOMs (upper part) form the set A, while variables of the set A
  and those with lower FOMs (lower part) form the set B.}
\label{tab:varnew}
\end{center}
\end{table}

\end{document}